\documentclass[usenatbib]{mn2e}
\usepackage{amsmath,fleqn}

\usepackage{epsf}

\newcommand\bF{{\bmath F}}

\newcommand\bU{{\bmath U}}
\newcommand\bu{{\bmath u}}
\newcommand\bv{{\bmath v}}

\newcommand\real{\mathrm{Re}}

\newcommand\half{{\textstyle\frac{1}{2}}}
\newcommand\third{{\textstyle\frac{1}{3}}}
\newcommand\twothirds{{\textstyle\frac{2}{3}}}

\newcommand\rmb{\mathrm{b}}

\newcommand\rmd{\mathrm{d}}
\newcommand\rme{\mathrm{e}}

\newcommand\rmi{\mathrm{i}}

\newcommand\rms{\mathrm{s}}

\newcommand\rmD{\mathrm{D}}
\newcommand\rmT{\mathrm{T}}
\newcommand\f{\frac}
\newcommand\p{\partial}
\newcommand\cst{\mathrm{constant}}

\title[Hydrodynamic instability in warped discs]
{Hydrodynamic instability in warped astrophysical discs}

\author[Gordon I.\ Ogilvie and Henrik N.\ Latter]
{Gordon I.\ Ogilvie and Henrik N.\ Latter\\
Department of Applied Mathematics and Theoretical Physics,
University of Cambridge, Centre for Mathematical Sciences,\\
Wilberforce Road, Cambridge CB3 0WA}

\begin{document}

\maketitle

\label{firstpage}
 
\begin{abstract}
  Warped astrophysical discs are usually treated as laminar viscous
  flows, which have anomalous properties when the disc is nearly
  Keplerian and the viscosity is small: fast horizontal shearing
  motions and large torques are generated, which cause the warp to
  evolve rapidly, in some cases at a rate that is inversely
  proportional to the viscosity. However, these flows are often
  subject to a linear hydrodynamic instability, which may produce
  small-scale turbulence and modify the large-scale dynamics of the
  disc.  We use a warped shearing sheet to compute the oscillatory
  laminar flows in a warped disc and to analyse their linear stability
  by the Floquet method.  We find widespread hydrodynamic instability
  deriving from the parametric resonance of inertial waves.  Even very
  small, unobservable warps in nearly Keplerian discs of low viscosity
  can be expected to generate hydrodynamic turbulence, or at least
  wave activity, by this mechanism.
\end{abstract}


\section{Introduction}

In the companion paper \citep[][hereafter Paper~I]{OL13}, we have
revisited the basic theory of warped astrophysical discs, in which the
orbital plane varies with radius. We have introduced a local model, a
warped shearing sheet, that is suitable for detailed analytical and
numerical studies of warped discs.  As well as rederiving the
nonlinear theory of laminar viscous discs \citep{1999MNRAS.304..557O}
by a simpler route, we have shown how the local model can be used more
generally to compute the internal torque that drives the large-scale
evolution of the shape and mass distribution of the disc, thereby
connecting the local and global dynamics.

An important feature of a warped disc is that the orbiting fluid
experiences a horizontal pressure gradient, antisymmetric about the
midplane, that oscillates at the orbital frequency (or nearly so, if
the shape of the disc is slowly varying in a non-rotating frame of
reference).  This force generates horizontal shearing motions, which
are resonantly amplified if the disc is Keplerian (or nearly so),
owing to the coincidence of the orbital and epicyclic frequencies.  By
transporting angular momentum in the radial direction, these motions
provide an internal torque that causes an anomalously rapid evolution
of the disc.

Existing treatments of warped discs represent these internal motions,
if at all, as laminar flows.  In \citet{1983MNRAS.202.1181P} their
amplitude is limited only by viscous damping, with the result that the
internal torque, and the rate of evolution of the warp, are inversely
proportional to the viscosity.  (In this context, viscosity is usually
taken to represent the effects of unresolved physical processes such
as small-scale turbulence.)  In a propagating bending wave
\citep{1995ApJ...438..841P} the time-dependence ensures that the
internal motions have a finite amplitude, even in the absence of
viscosity, but the linear theory often predicts hypersonic flows above
and below the midplane.  The same is true in the nonlinear theory of
\citet{1999MNRAS.304..557O}, which we have revisited in Paper~I.

It has previously been noted that these oscillatory shear flows are
likely to be linearly unstable
\citep{1995MNRAS.274..987P,2000MNRAS.318.1005G}.  To the best of our
knowledge, instability has not been detected in any of the global
numerical simulations of warped discs (many of which are cited in
Paper~I), probably because they have insufficient spatial resolution
or too much viscosity, either explicit or numerical, to allow the
unstable modes to develop.  Therefore these simulations also represent
warped discs as laminar flows and may be missing an important physical
process.

Hydrodynamic instability is likely to lead to turbulence, or at least
significant wave activity, and to alter the internal flows in a warped
disc.  We expect an important modification of the large-scale dynamics
of warped discs, and this is one reason why we are motivated to make a
detailed study of the instability.  Another reason is that it will
provide a source of hydrodynamic activity in Keplerian discs,
especially in poorly ionized regions where the magnetorotational
instability (MRI) is inoperative.  These secondary motions could be
important in determining many of the physical properties of these
regions, for example by stirring up the dead zones of protoplanetary
discs.  Note that only a very small warp may be required to excite
these motions, owing to the resonant amplification of the internal
flows in a Keplerian disc.  Such small but dynamically significant
warps may not be directly observable even with high-resolution
imaging.  More generally, we are interested in the interplay between
hydrodynamic instability resulting from a warp and other sources of
turbulence and transport in astrophysical discs, including the MRI.

The likelihood of a local linear hydrodynamic instability in warped
discs was noted by \citet{1995MNRAS.274..987P}, and the first explicit
calculations were made by \citet{2000MNRAS.318.1005G}.\footnote{The
  problem considered by \citet{1993MNRAS.260..323K}, although
  motivated by warped discs, is too simplistic to be directly
  applicable.}  \citet{2000MNRAS.318.1005G} did not represent the warp
itself, but considered the horizontal shearing motions, oscillating at
the local orbital frequency, that are driven by the warp.  These
motions are linearly unstable when the associated shear rate is
sufficiently large compared to the viscosity of the fluid, the
instability taking the form of a parametric resonance of inertial
waves.  In nonlinear local simulations, the shearing motions decay as
they lose energy to inertial waves.  \citet{2000MNRAS.318...47T}
studied the decay of similar flows in the presence of
magnetohydrodynamic (MHD) turbulence driven by the MRI.

An effect related to the parametric instability of warped discs is the
mode-coupling process investigated by \citet{2004PASJ...56..905K} and
\citet{2008MNRAS.386.2297F}.  This allows a warp in the inner part of
a disc around a black hole to excite inertial waves that are trapped
near the radius at which the epicyclic frequency is maximal, and which
may be related to observed high-frequency quasi-periodic oscillations
from accreting black holes.

In the local hydrodynamic and MHD simulations of
\citet{2000MNRAS.318.1005G} and \citet{2000MNRAS.318...47T}, the
horizontal shearing motions were imposed as initial conditions and
were allowed to decay through instability or interaction with
turbulence.  However, in order to reach a better understanding of the
role of instabilities or turbulence in the dynamics of warped discs, a
model is required in which the effect of the warp can be maintained
and the dynamics can reach a quasi-steady state.  Such a model is
provided by the warped shearing sheet defined in Paper~I.

To prepare the way for future nonlinear simulations in the warped
shearing sheet, we tackle the linear problem in this paper, analysing
the hydrodynamic stability of the laminar flows computed in Paper~I.
Our analysis is closely related to that of
\citet{2000MNRAS.318.1005G}, but is more detailed and is based on the
nonlinear laminar flow solutions computed in a warped shearing sheet.
We compute the growth rates of the unstable modes and their variation
with several parameters.  In certain regimes the instability can be
clearly identified as a parametric resonance of inertial waves, in
agreement with previous work.  We also speculate on the nonlinear
outcome and its consequences for warped discs, as well as making
suggestions for further investigation.

\section{Local model and laminar flows}

In this section we summarize the equations of Paper~I on which our
analysis is based.  The local model of a warped disc introduced in
that paper is based on a circular reference orbit of radius $r_0$ and
angular velocity $\Omega_0$, where the dimensionless rate of orbital
shear is $q=-\rmd\ln\Omega/\rmd\ln r$, equal to ${\textstyle\f{3}{2}}$
for Keplerian orbits.  The initial construction is identical to the
standard shearing sheet or box and leads to the hydrodynamic equations
\begin{equation}
  \rmD u_x-2\Omega_0u_y=2q\Omega_0^2x-\p_xh,
\end{equation}
\begin{equation}
  \rmD u_y+2\Omega_0u_x=-\p_yh,
\end{equation}
\begin{equation}
  \rmD u_z=-\Omega_0^2z-\p_zh,
\end{equation}
\begin{equation}
  \rmD h=-c_\rms^2(\p_xu_x+\p_yu_y+\p_zu_z),
\end{equation}
where
\begin{equation}
  \rmD=\p_t+u_x\p_x+u_y\p_y+u_z\p_z
\label{d}
\end{equation}
is the Lagrangian derivative and $\bu$ is the velocity.  For
simplicity, we consider an isothermal gas for which
\begin{equation}
  p=c_\rms^2\rho,
\end{equation}
where $c_\rms=\cst$ is the isothermal sound speed, and we make use of
the pseudo-enthalpy
\begin{equation}
  h=c_\rms^2\ln\rho+\cst.
\end{equation}

To represent a warped disc we introduce the warped shearing
coordinates
\begin{equation}
  t'=t,
\end{equation}
\begin{equation}
  x'=x,
\label{xp}
\end{equation}
\begin{equation}
  y'=y+q\tau x,
\label{yp}
\end{equation}
\begin{equation}
  z'=z+|\psi|\cos\tau\,x,
\end{equation}
where
\begin{equation}
  \tau=\Omega_0t'=\Omega_0t
\end{equation}
is the orbital phase and $|\psi|$ is the dimensionless warp amplitude.
These coordinates follow the warped orbital motion.  Partial
derivatives transform according to
\begin{equation}
  \p_t=\p_t'+q\Omega_0x\p_y'-|\psi|\Omega_0\sin\tau\,x\p_z',
\end{equation}
\begin{equation}
  \p_x=\p_x'+q\tau\,\p_y'+|\psi|\cos\tau\,\p_z',
\end{equation}
\begin{equation}
  \p_y=\p_y',
\end{equation}
\begin{equation}
  \p_z=\p_z',
\end{equation}
so that
\begin{equation}
  \rmD=\p_t'+v_x\p_x'+(v_y+q\tau v_x)\p_y'+(v_z+|\psi|\cos\tau\,v_x)\p_z',
\end{equation}
where
\begin{equation}
  v_x=u_x,
\end{equation}
\begin{equation}
  v_y=u_y+q\Omega_0x,
\end{equation}
\begin{equation}
  v_z=u_z-|\psi|\Omega_0\sin\tau\,x
\end{equation}
are the relative velocity components.

The hydrodynamic equations are therefore transformed into
\begin{eqnarray}
  \rmD v_x-2\Omega_0v_y=-(\p_x'+q\tau\,\p_y'+|\psi|\cos\tau\,\p_z')h,
\label{dvx}
\end{eqnarray}
\begin{equation}
  \rmD v_y+(2-q)\Omega_0v_x=-\p_y'h,
\end{equation}
\begin{equation}
  \rmD v_z+|\psi|\Omega_0\sin\tau\,v_x=-\Omega_0^2z'-\p_z'h,
\label{dvz}
\end{equation}
\begin{equation}
  \rmD h=-c_\rms^2[(\p_x'+q\tau\,\p_y'+|\psi|\cos\tau\,\p_z')v_x+\p_y'v_y+\p_z'v_z].
\label{dh}
\end{equation}
These equations are easily extended to include a dynamic shear
viscosity $\mu=\alpha p/\Omega_0$ and a dynamic bulk viscosity
$\mu_\rmb=\alpha_\rmb p/\Omega_0$, where $\alpha$ and $\alpha_\rmb$
are constant dimensionless coefficients.  We do not write out the
viscous terms explicitly, however.

The simplest solutions of equations (\ref{dvx})--(\ref{dh}), extended
to include viscosity, are laminar flows of the form
\begin{equation}
  v_x(z',t')=u(\tau)\Omega_0z',
\label{vx}
\end{equation}
\begin{equation}
  v_y(z',t')=v(\tau)\Omega_0z',
\end{equation}
\begin{equation}
  v_z(z',t')=w(\tau)\Omega_0z',
\end{equation}
\begin{equation}
  h(z',t')=c_\rms^2f(\tau)-\half\Omega_0^2z'^2g(\tau),
\label{h}
\end{equation}
where $u$, $v$, $w$, $f$ and~$g$ are dimensionless $2\pi$-periodic
functions that satisfy the equations
\begin{eqnarray}
  \lefteqn{\rmd_\tau u+(w+|\psi|\cos\tau\,u)u-2v=|\psi|\cos\tau\,g}&\nonumber\\
  &&-(\alpha_\rmb+\third\alpha)|\psi|\cos\tau\,g(w+|\psi|\cos\tau\,u)\nonumber\\
  &&-\alpha g[|\psi|\sin\tau+(1+|\psi|^2\cos^2\tau)u],
\label{du}
\end{eqnarray}
\begin{eqnarray}
  \lefteqn{\rmd_\tau v+(w+|\psi|\cos\tau\,u)v+(2-q)u}&\nonumber\\
  &&=-\alpha g[-q|\psi|\cos\tau+(1+|\psi|^2\cos^2\tau)v],
\end{eqnarray}
\begin{eqnarray}
  \lefteqn{\rmd_\tau w+(w+|\psi|\cos\tau\,u)w+|\psi|\sin\tau\,u=g-1}&\nonumber\\
  &&-(\alpha_\rmb+\third\alpha)g(w+|\psi|\cos\tau\,u)\nonumber\\
  &&-\alpha g[|\psi|^2\sin\tau\cos\tau+(1+|\psi|^2\cos^2\tau)w],
\end{eqnarray}
\begin{equation}
  \rmd_\tau f=-(w+|\psi|\cos\tau\,u),
\label{df}
\end{equation}
\begin{equation}
  \rmd_\tau g=-2(w+|\psi|\cos\tau\,u)g,
\label{dg}
\end{equation}
where $\rmd_\tau$ denotes the ordinary derivative $\rmd/\rmd\tau$.
Numerical solutions of these equations are presented in Paper~I.

\section{Stability of laminar flows}
\label{s:stability}

\subsection{Simple expectations}

Before embarking on a detailed analysis of the stability of the
laminar flows, we make some very simple estimates, omitting factors of
order unity.

We are usually interested in discs that are nearly Keplerian
($|q-{\textstyle\f{3}{2}}|\ll1$) and of low viscosity ($\alpha\ll1$).
In such discs the horizontal laminar flows are resonant and can reach
large amplitudes, even if the warp amplitude $|\psi|$ is small.

The magnitude of~$u$ (or $v$, which is similar) can be simply
estimated as follows.  The horizontal motion is a linear oscillator,
whose natural frequency is the epicyclic frequency.  It is forced at
the orbital frequency by the horizontal pressure gradients associated
with the warp, and is damped by viscosity.  The problem is similar to
a damped harmonic oscillator forced at, or close to, its natural
frequency; the response of the oscillator is limited either by
detuning or damping, whichever is larger.  When detuning of the
resonance due to non-Keplerian rotation is the limiting factor, $u$
scales as $|\psi|/|q-{\textstyle\f{3}{2}}|$ (valid when this quantity
is $\la1$).  When viscous damping of the horizontal flows is the
limiting factor, $u$ scales as $|\psi|/\alpha$ (valid when this
quantity is $\la1$).  Examples of these scalings can be seen in figs~6
and~7 of Paper~I.

These oscillatory shear flows are unstable in the absence of
viscosity, and the inviscid growth rate of the instability scales with
the shear rate $u\Omega$ \citep{2000MNRAS.318.1005G}.  Since the
unstable modes have a non-trivial vertical structure and a horizontal
wavelength related to the scale-height of the disc, $H$, viscous
damping quenches the instability if $\nu/H^2\ga u\Omega$, i.e.\ if
$\alpha\ga u$.  Therefore (omitting factors of order unity)
instability is expected when $|\psi|\ga|q-{\textstyle\f{3}{2}}|\alpha$
or $\alpha^2$, whichever is larger.  These are only very rough
estimates or scaling relations; detailed results are obtained in
Section~\ref{s:numerical_stability} below.

\subsection{Linearized equations}

To examine the stability of the laminar flows we linearize the basic
equations (\ref{dvx})--(\ref{dh}) about the solution
(\ref{vx})--(\ref{h}), introducing infinitesimal perturbations of the
form
\begin{equation}
  \real[\delta\bv(z',\tau)\exp(\rmi k_x'x'+\rmi k_y'y')],
\end{equation}
etc., where $(k_x',k_y')$ is a real horizontal wavevector.  The
linearized equations admit solutions of this plane-wave form because
the basic equations (\ref{dvx})--(\ref{dh}) from which they are
derived do not contain $x'$ or $y'$ explicitly, and the basic state
about which we are linearizing is also independent of~$x'$ and~$y'$.
Using the relations~(\ref{xp}) and~(\ref{yp}) between the primed and
unprimed horizontal coordinates, we find
\begin{equation}
  \exp(\rmi k_x'x'+\rmi k_y'y')=\exp(\rmi k_xx+\rmi k_yy),
\end{equation}
where
\begin{equation}
  (k_x,k_y)=(k_x'+qk_y'\tau,k_y')
\end{equation}
is a wavevector that may depend on time.  If $k_y\ne0$ then $k_x$
depends linearly on $\tau$ and we have a shearing wave, which
corresponds to a non-axisymmetric spiral disturbance in the global
geometry.  If $k_y=0$ then $k_x=k_x'$ is a constant: the wave is
axisymmetric and does not shear.

We also adopt units in which $\Omega_0=1$ and $c_\rms=1$, meaning that
the unit of length is the scale-height $c_\rms/\Omega_0$ of an
unwarped disc and the unit of time is $\Omega_0^{-1}$.  The linearized
equations in the absence of viscosity then take the form
\begin{eqnarray}
  \lefteqn{\mathcal{D}\delta v_x+(\delta v_z+|\psi|\cos\tau\,\delta v_x)u-2\,\delta v_y}&\nonumber\\
  &&=-(\rmi k_x+|\psi|\cos\tau\,\p_z')\delta h,
\label{caldvx}
\end{eqnarray}
\begin{equation}
  \mathcal{D}\delta v_y+(\delta v_z+|\psi|\cos\tau\,\delta v_x)v+(2-q)\delta v_x=-\rmi k_y\delta h,
\end{equation}
\begin{equation}
  \mathcal{D}\delta v_z+(\delta v_z+|\psi|\cos\tau\,\delta v_x)w+|\psi|\sin\tau\,\delta v_x=-\p_z'\delta h,
\end{equation}
\begin{eqnarray}
  \lefteqn{\mathcal{D}\delta h-(\delta v_z+|\psi|\cos\tau\,\delta v_x)gz'=-\rmi k_x\delta v_x}\nonumber\\
  &&-|\psi|\cos\tau\,\p_z'\delta v_x-\rmi k_y\delta v_y-\p_z'\delta v_z,
\label{caldh}
\end{eqnarray}
where
\begin{equation}
  \mathcal{D}=\p_\tau+\rmi(k_xu+k_yv)z'+(w+|\psi|\cos\tau\,u)z'\p_z'.
\label{cald}
\end{equation}
For an axisymmetric wave ($k_y=0$), $k_x$ is constant and
the coefficients of the linearized equations have a periodic
dependence on $\tau$, with period $2\pi$.  For a shearing wave
($k_y\ne0$), the coefficients are non-periodic because $k_x$ depends
linearly on $\tau$.

The associated equation for the energy of the perturbation is
(assuming suitable boundary conditions in $z$)
\begin{eqnarray}
  \lefteqn{\rmd_\tau\int\rho\left(\half|\delta\bv|^2+\half|\delta h|^2\right)\rmd z'}&\nonumber\\
  &&=-\real\int\rho\bigg[|\delta v_x|^2|\psi|\cos\tau\,u+\delta v_x^*\delta v_y(-q+|\psi|\cos\tau\,v)\nonumber\\
  &&\qquad\qquad\qquad+\delta v_x^*\delta v_z(|\psi|\sin\tau+|\psi|\cos\tau\,w+u)\nonumber\\
  &&\qquad\qquad\qquad+\delta v_y^*\delta v_zv+|\delta v_z|^2w\bigg]\rmd z'.
\label{perturbation_energy}
\end{eqnarray}
The right-hand side represents the exchange of energy with the laminar
flow and the orbital shear through Reynolds stresses.  This equation
can be used to bound the growth rate of the perturbation, as discussed
in Appendix~\ref{s:bound}.

\subsection{Projection on to orthogonal polynomials}

To solve equations (\ref{caldvx})--(\ref{cald}) we use a Galerkin
spectral method, projecting the equations on to a basis of Hermite
polynomials
\begin{equation}
  \mathrm{He}_n(x)=\rme^{x^2/2}\left(-\f{\rmd}{\rmd x}\right)^n\rme^{-x^2/2},\qquad
  n=0,1,2,\dots
\end{equation}
as used by \citet{1987PASJ...39..457O} and many other authors:
\begin{equation}
  \delta v_x(z',\tau)=\sum_{n=0}^\infty u_n(\tau)\,\mathrm{He}_n(z'),
\end{equation}
\begin{equation}
  \delta v_y(z',\tau)=\sum_{n=0}^\infty v_n(\tau)\,\mathrm{He}_n(z'),
\end{equation}
\begin{equation}
  \delta v_z(z',\tau)=\sum_{n=1}^\infty w_n(\tau)\,\mathrm{He}_{n-1}(z'),
\end{equation}
\begin{equation}
  \delta h(z',\tau)=\sum_{n=0}^\infty h_n(\tau)\,\mathrm{He}_n(z').
\end{equation}
We further define $u_n=v_n=h_n=0$ for $n<0$ and $w_n=0$ for $n<1$.
The Hermite polynomials, familiar from the quantum harmonic
oscillator, are particularly well suited to linear waves in isothermal
discs, having a Gaussian weight function that is proportional to the
density of an unwarped isothermal disc.  Using the relations
\begin{equation}
  \mathrm{He}_n'(x)=n\,\mathrm{He}_{n-1}(x),
\end{equation}
\begin{equation}
  x\,\mathrm{He}_n(x)=\mathrm{He}_{n+1}(x)+n\,\mathrm{He}_{n-1}(x),
\label{xhenx}
\end{equation}
we find
\begin{eqnarray}
  \lefteqn{\rmd_\tau u_n+\rmi(k_xu+k_yv)[u_{n-1}+(n+1)u_{n+1}]}&\nonumber\\
  &&+(w+|\psi|\cos\tau\,u)[nu_n+(n+1)(n+2)u_{n+2}]\nonumber\\
  &&+u(w_{n+1}+|\psi|\cos\tau\,u_n)-2v_n=-\rmi k_xh_n\nonumber\\
  &&-|\psi|\cos\tau\,(n+1)h_{n+1},
\label{dun}
\end{eqnarray}
\begin{eqnarray}
  \lefteqn{\rmd_\tau v_n+\rmi(k_xu+k_yv)[v_{n-1}+(n+1)v_{n+1}]}&\nonumber\\
  &&+(w+|\psi|\cos\tau\,u)[nv_n+(n+1)(n+2)v_{n+2}]\nonumber\\
  &&+v(w_{n+1}+|\psi|\cos\tau\,u_n)+(2-q)u_n=-\rmi k_yh_n,
\label{dvn}
\end{eqnarray}
\begin{eqnarray}
  \lefteqn{\rmd_\tau w_n+\rmi(k_xu+k_yv)(w_{n-1}+nw_{n+1})}&\nonumber\\
  &&+(w+|\psi|\cos\tau\,u)[(n-1)w_n+n(n+1)w_{n+2}]\nonumber\\
  &&+w(w_n+|\psi|\cos\tau\,u_{n-1})+|\psi|\sin\tau\,u_{n-1}=-nh_n,
\label{dwn}
\end{eqnarray}
\begin{eqnarray}
  \lefteqn{\rmd_\tau h_n+\rmi(k_xu+k_yv)[h_{n-1}+(n+1)h_{n+1}]}&\nonumber\\
  &&+(w+|\psi|\cos\tau\,u)[nh_n+(n+1)(n+2)h_{n+2}]\nonumber\\
  &&-(g-1)\{w_n+(n+1)w_{n+2}\nonumber\\
  &&\qquad+|\psi|\cos\tau[u_{n-1}+(n+1)u_{n+1}]\}\nonumber\\
  &&=-\rmi k_xu_n+|\psi|\cos\tau\,u_{n-1}-\rmi k_yv_n+w_n.
\label{dhn}
\end{eqnarray}
Equations (\ref{dun}), (\ref{dvn}) and (\ref{dhn}) are to be solved
for $n\ge0$, and equation~(\ref{dwn}) for $n\ge1$.  The equivalent
equations including viscosity are given in Appendix~\ref{s:viscous}.

\subsection{Solutions in the absence of a warp}
\label{s:absence}

\begin{figure*}
\centerline{\epsfysize8cm\epsfbox{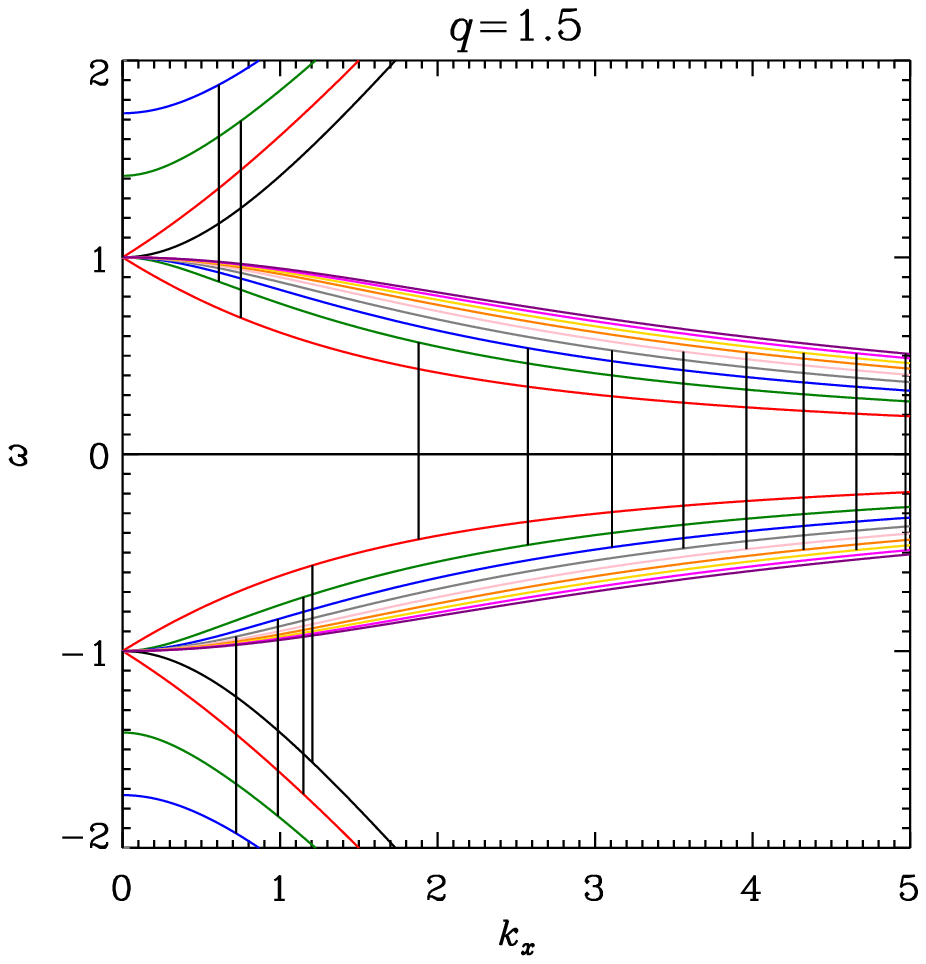}\quad\epsfysize8cm\epsfbox{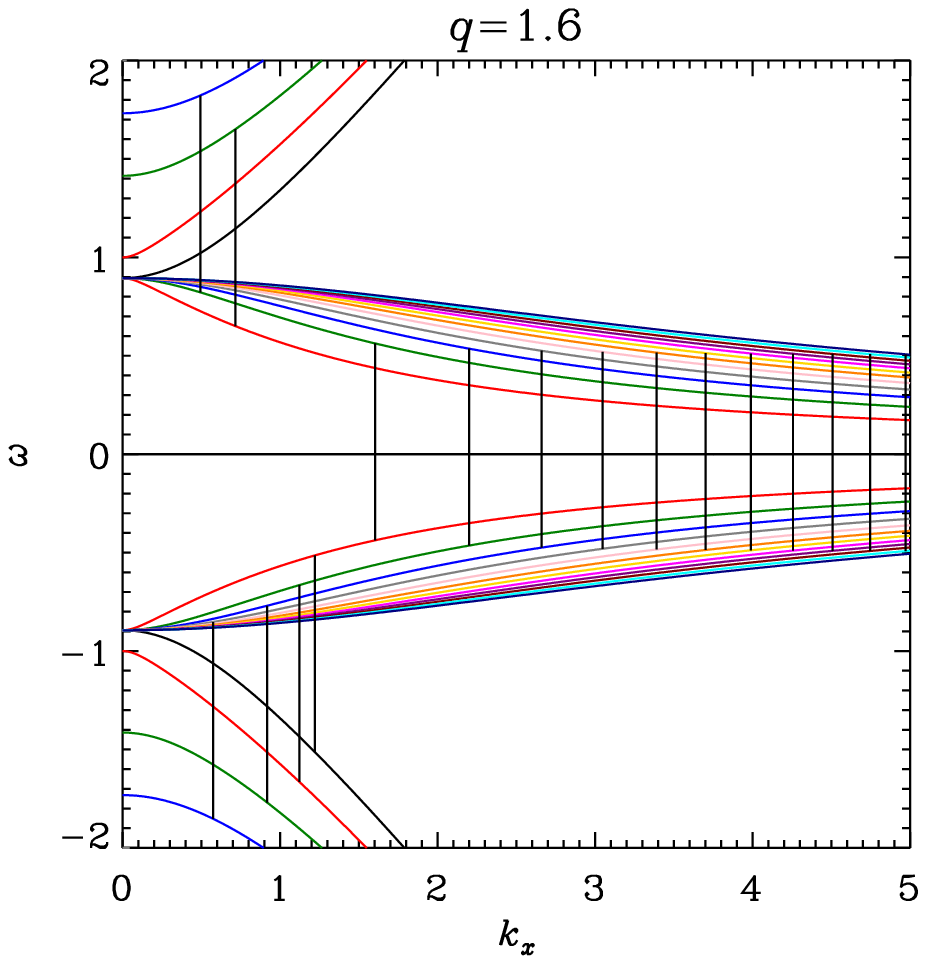}}
\caption{Dispersion relation for axisymmetric waves ($k_y=0$) in an
  unwarped disc in the absence of viscosity, for $q=1.5$
  (\textit{left}) and $q=1.6$ (\textit{right}).  The angular frequency
  $\omega$ of the waves is in units of~$\Omega_0$, and the radial
  wavenumber $k_x$ is in units of $\Omega_0/c_\rms$.  Different
  colours indicate different values of the vertical mode number $n$
  (0: black, 1: red, 2: green, 3: blue, 4: grey, 5: pink, 6: orange,
  7: gold, 8: magenta, 9: purple, 10: maroon, 11: cyan, 12: navy);
  only modes with $n\le9$ are shown for $q=1.5$.  Vertical bars of
  length~$1$ show the possible resonant couplings between neighbouring
  modes.  These occur at wavenumbers for which two waves exist with
  vertical mode numbers $n$ and $n+1$ and with angular frequencies
  $\omega$ and $\omega+1$.}
\label{f:resonances}
\end{figure*}

In the absence of a warp, $|\psi|=u=v=w=g-1=0$ and the inviscid
equations reduce to
\begin{equation}
  \rmd_\tau u_n-2v_n=-\rmi k_xh_n,
\end{equation}
\begin{equation}
  \rmd_\tau v_n+(2-q)u_n=-\rmi k_yh_n,
\end{equation}
\begin{equation}
  \rmd_\tau w_n=-nh_n,
\end{equation}
\begin{equation}
  \rmd_\tau h_n=-\rmi k_xu_n-\rmi k_yv_n+w_n,
\end{equation}
with no coupling between different values of~$n$.  The same property
holds when viscosity is included, but only if
$\alpha_\rmb=\twothirds\alpha$.  For axisymmetric waves ($k_y=0$),
solutions $\propto\rme^{-\rmi\omega\tau}$ exist, with angular
frequency $\omega$ given by, in the inviscid case,
\begin{equation}
  (-\omega^2+n)[-\omega^2+2(2-q)]-k_x^2\omega^2=0,
\label{dr}
\end{equation}
\citep[cf.][]{1987PASJ...39..457O} and, in the viscous case, a more
complicated quartic equation.  For $n\ge1$, the high-frequency
solutions of equation~(\ref{dr}) have an acoustic character, while the
low-frequency solutions have an inertial character.  In the case
$n=0$, however, the high-frequency solution is the well known
two-dimensional (acoustic--inertial) wave, while the low-frequency
solution is a zero-frequency vortical mode.  The dispersion relation
is illustrated in Fig.~\ref{f:resonances}.

\subsection{Solutions in the presence of a warp}
\label{s:presence}

In the presence of a warp, the different Hermite modes are coupled by
numerous terms in equations (\ref{dun})--(\ref{dhn}).  For example,
the terms proportional to $\rmi(k_xu+k_yv)$, which derive from the
shearing of the waves by the laminar flow, couple neighbouring modes
in such a way that an infinite ladder is produced.  Indeed, the
linearized equations (\ref{caldvx})--(\ref{caldh}) have no solutions
that are strictly polynomial in $z'$ in the presence of a warp,
because the $\rmi(k_xu+k_yv)z'$ term in $\mathcal{D}$ generates an
infinite power series in $z'$.  Another type of coupling is provided
by the terms proportional to $(g-1)$ in equation~(\ref{dhn}).  These
terms arise because the scale-height of the warped disc is
time-dependent and different from that of the unwarped disc on which
the Hermite polynomials are based.  For computational purposes, we
truncate the system at some order $N$ by setting $u_n$, $v_n$, $w_n$
and $h_n$ to zero for $n\ge N$.

The parametric instability, at least in its simplest form, involves a
three-wave coupling between the warp (together with the associated
laminar flow) and two inertial waves \citep{2000MNRAS.318.1005G}.  The
theory is worked out in detail in Appendix~\ref{s:parametric}, which
is based on an expansion of the equations for small $|\psi|$.  In the
local model, the warp and its associated laminar flow have radial
wavenumber $0$, vertical mode number $1$ and frequency $1$.  In order
to couple to the warp, the inertial waves should therefore have the
same radial wavenumber as each other, while their vertical mode
numbers should differ by $1$ (owing to equation~\ref{xhenx}) and their
frequencies should add up to (or differ by) $1$.  The density of the
spectrum of inertial waves means that there are infinitely many
wavenumbers at which this coupling is possible; several of these are
illustrated in Fig.~\ref{f:resonances}.  (Couplings of acoustic and
inertial waves are found not to lead to instability.)

We investigate only axisymmetric waves ($k_y=0$) in this paper,
looking for instabilities of the laminar flows.  As noted above, in
this case the linearized equations have coefficients that are
$2\pi$-periodic in $\tau$.  Our truncated system is a set of $4N-1$
ordinary differential equations (ODEs) and we analyse their solutions
using Floquet theory.  The characteristic solutions (eigenmodes) of
the system of ODEs are $2\pi$-periodic functions of~$\tau$ multiplied
by $\rme^{s\tau}$, where $s$ is a complex growth rate.  Choosing a set
of $4N-1$ linearly independent initial conditions (by setting all
variables except one to zero at $\tau=0$) and integrating the ODEs
from $\tau=0$ to $\tau=2\pi$ using a Runge--Kutta method with adaptive
stepsize, we evaluate the monodromy matrix that determines the
evolution of an arbitrary initial condition over one period.  We
calculate the eigenvalues of the monodromy matrix, which are the
characteristic multipliers $\rme^{2\pi s}$, and deduce the complex
growth rates.  Growing solutions are found when $\real(s)>0$, and
indicate linear instability of the laminar flow.  (The imaginary part
of~$s$ gives the angular frequency of the eigenmode, but this is only
defined modulo~$1$.)

This numerical method has been verified by reproducing the solutions
of the algebraic dispersion relation~(\ref{dr}) and its viscous
equivalent in the absence of a warp.  In the presence of a warp, as
described below, it produces results that agree with the analytical
calculations of parametric instability in Appendix~\ref{s:parametric}.
Equations (\ref{dun})--(\ref{dhn}) were also solved with a
pseudospectral method, which provides a separate numerical check on
the results. As noted above, the characteristic solutions are
$2\pi$-periodic functions of $\tau$ multiplied by $\rme^{s\tau}$, and
if this common factor is extracted from the solutions, a term
involving the complex growth rate $s$ appears in each equation.  After
truncating the number of Hermite modes to $N$, the $\tau$ domain is
partitioned into $M$ segments and the $\tau$ derivatives approximated
using Fourier collocation. This yields a square matrix of size $4MN$,
the eigenvalues of which are the complex growth rates $s$. Good
agreement was achieved with the monodromy approach described in the
previous paragraph.

\subsection{Numerical results}
\label{s:numerical_stability}

\begin{figure*}
\centerline{\epsfysize7.5cm\epsfbox{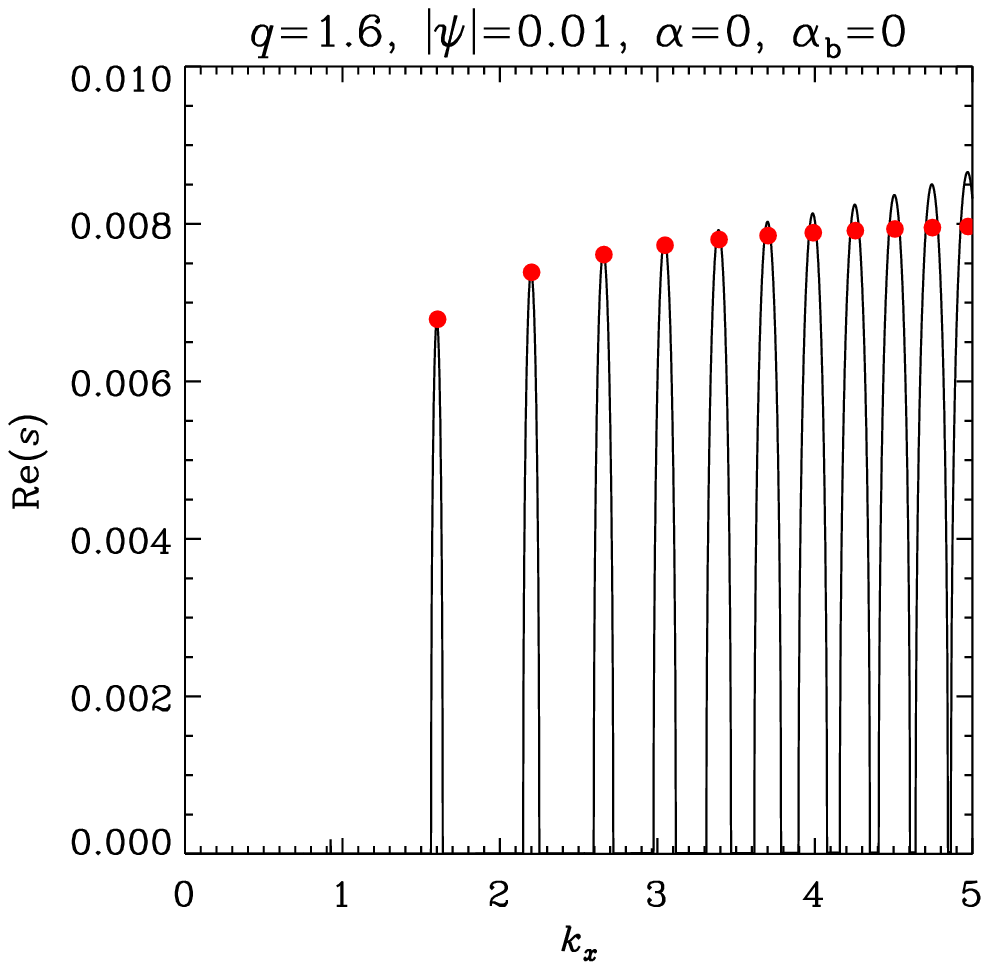}\qquad\epsfysize7.5cm\epsfbox{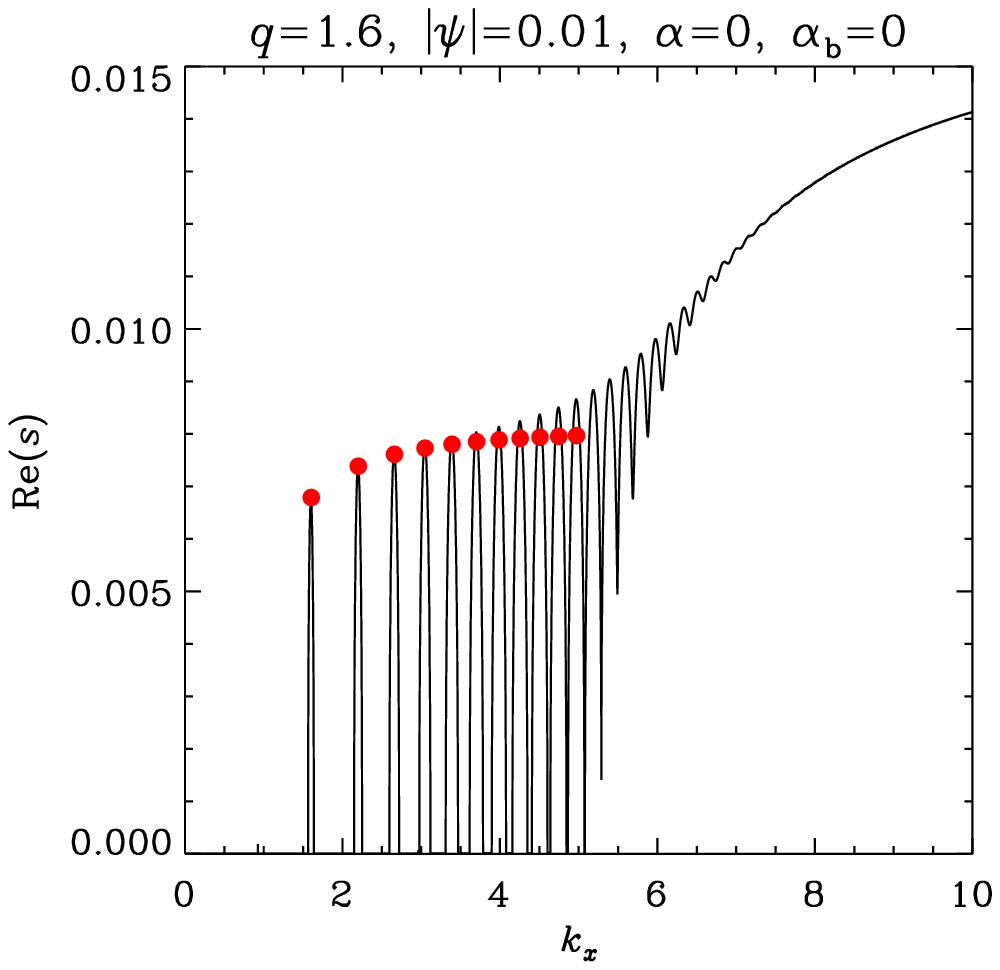}}
\centerline{\epsfysize7.5cm\epsfbox{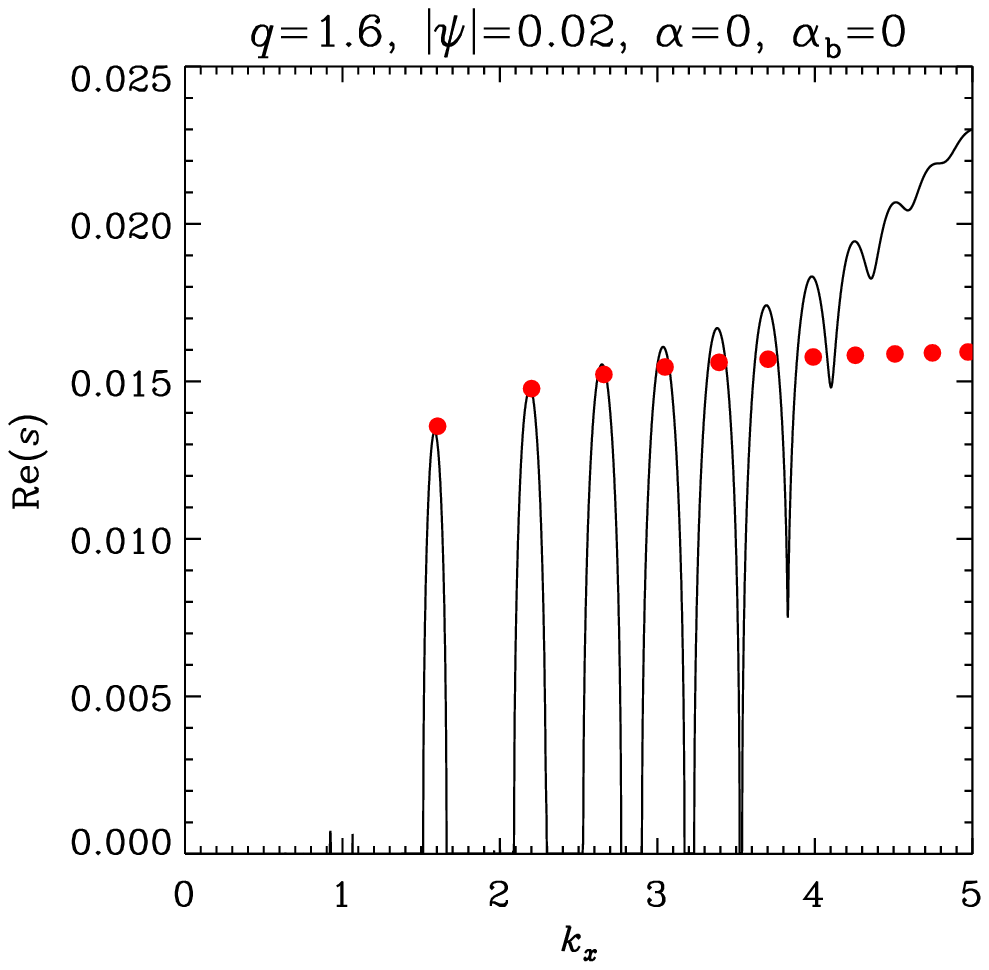}\qquad\epsfysize7.5cm\epsfbox{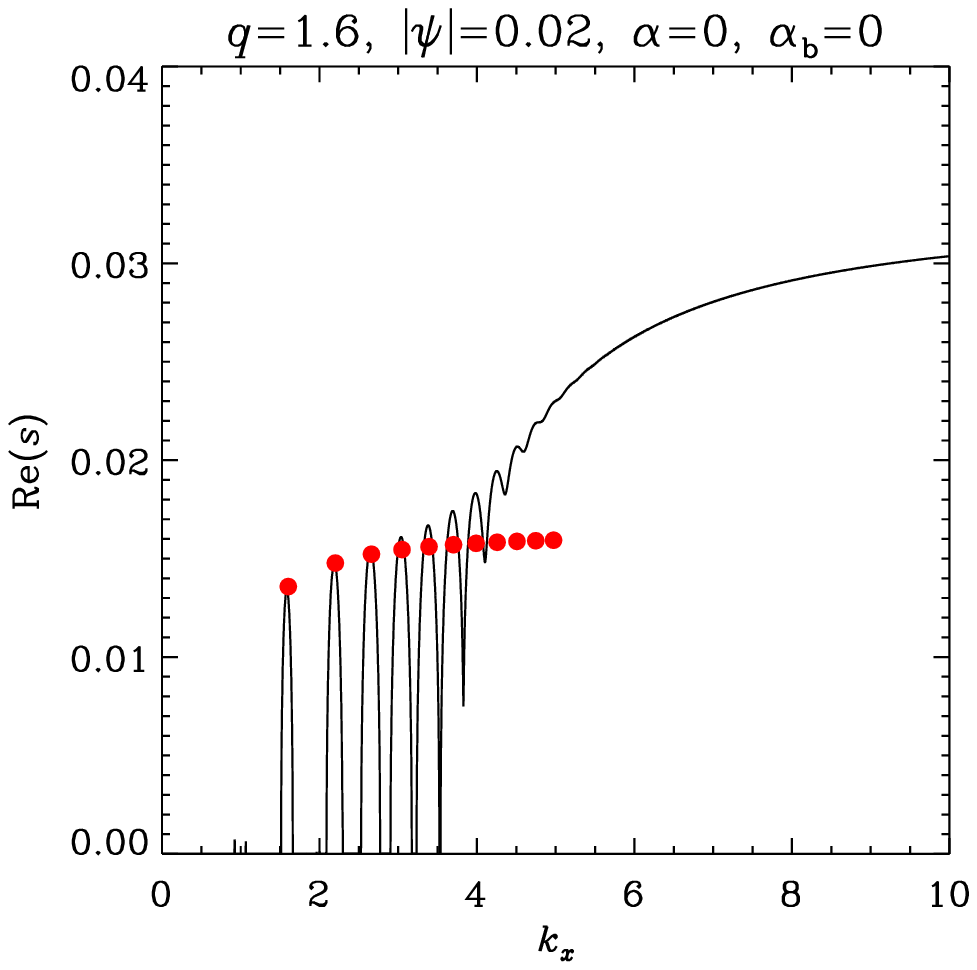}}
\centerline{\epsfysize7.5cm\epsfbox{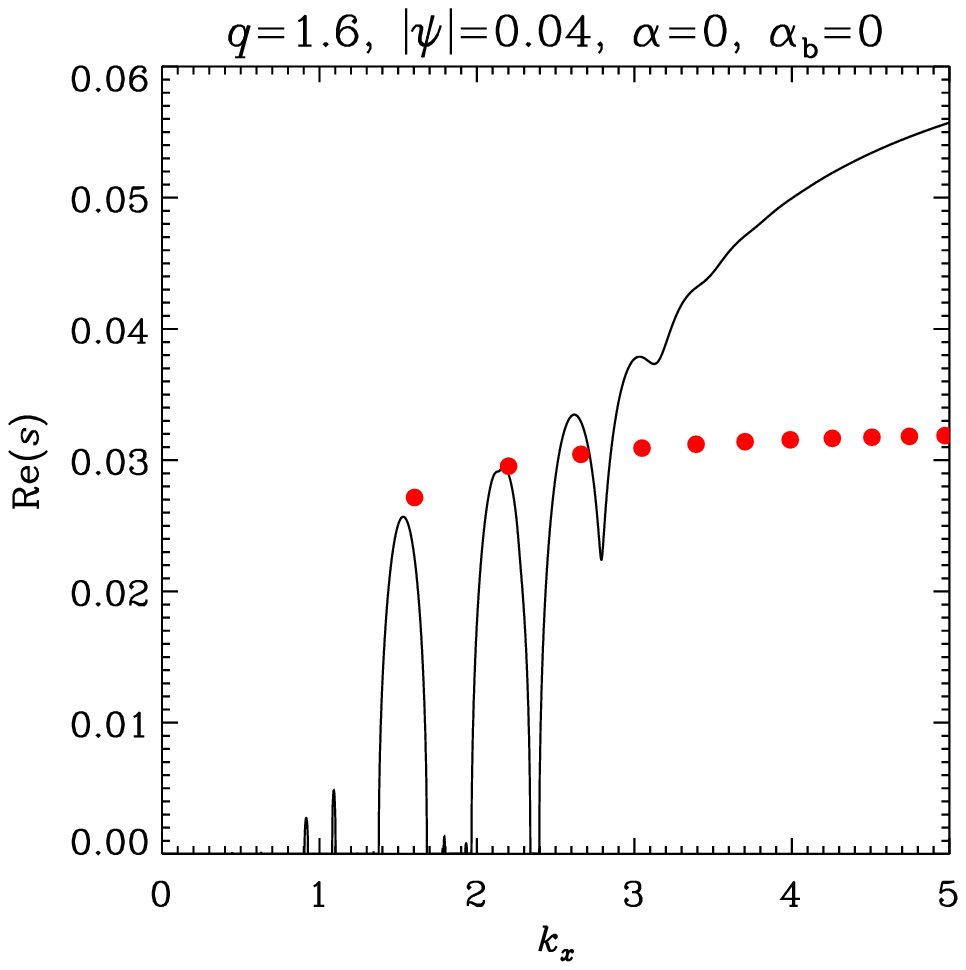}\qquad\epsfysize7.5cm\epsfbox{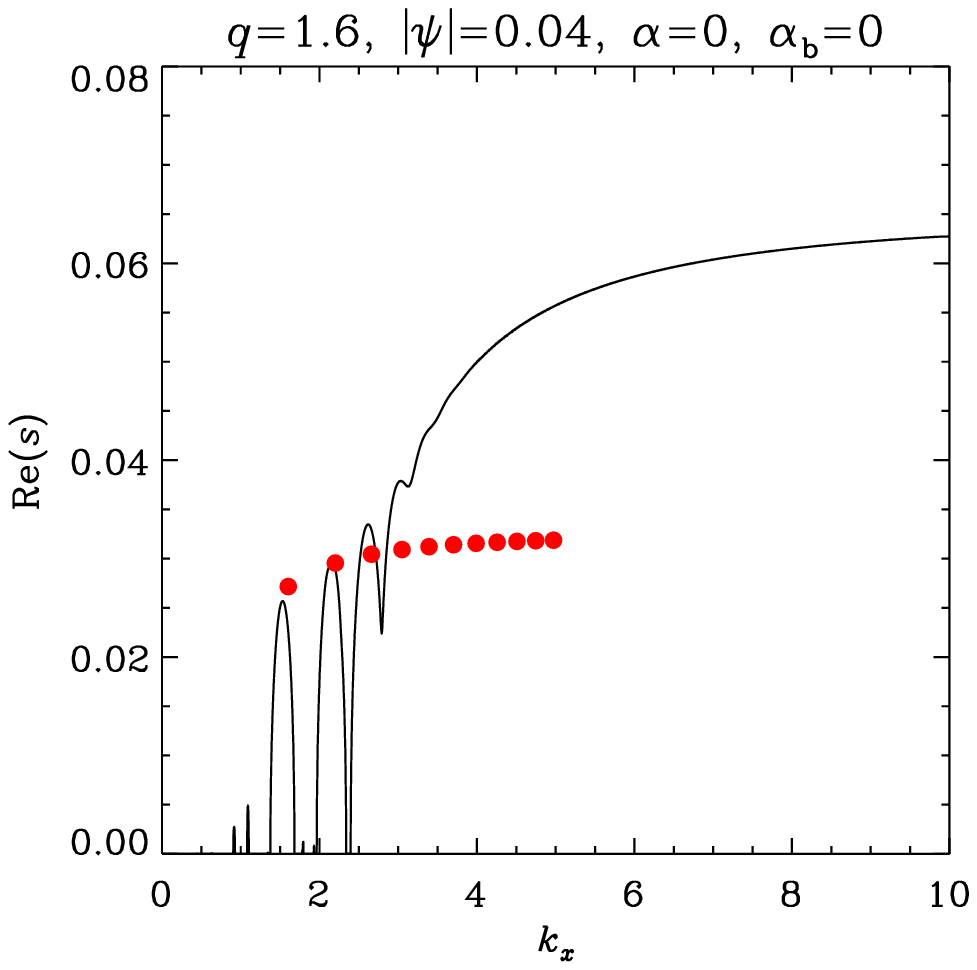}}
\caption{Linear stability of laminar flows for $q=1.6$.  The growth
  rate of the fastest growing mode (in units of~$\Omega_0$) is plotted
  versus the radial wavenumber (in units of $\Omega_0/c_\rms$).  The
  red points show the growth rate at the centre of each band of
  parametric instability (up to $k_x=5$), based on the analysis in
  Appendix~\ref{s:nonkeplerian}.  Larger growth rates are obtained
  where the resonance bands overlap.  The left panels require
  resolutions of up to $N=30$ for convergence.  The right panels,
  which illustrate the behaviour for larger wavenumbers, require up to
  $N=90$.}
\label{f:stability_q=1.6}
\end{figure*}

We begin a discussion of the numerical results by considering the
relatively simple case of an inviscid non-Keplerian disc with $q=1.6$
and a small warp of amplitude $|\psi|\ll1$.  We compute the growth
rate of the fastest growing mode for a range of radial wavenumbers and
plot the results in Fig.~\ref{f:stability_q=1.6}.  Instability occurs
in bands of~$k_x$ centred on the wavenumbers predicted by the analysis
of three-wave coupling; the red points in the figure show the growth
rate at the centre of each band (up to $k_x=5$) calculated in
Appendix~\ref{s:nonkeplerian}.  These growth rates are proportional to
$|\psi|$ and tend to a limit as $k_x\to\infty$.  The bands of
instability have a non-zero width, also proportional to $|\psi|$,
because the parametric instability permits some detuning.  The first
few bands are well separated, but eventually they overlap and merge,
producing a continuous curve.  The limiting growth rate in the
continuous regime is twice the limiting value predicted by a single
three-wave coupling, because each mode ($n$) can engage simultaneously
in three-wave couplings with its two neighbours ($n\pm1$)
\citep{2000MNRAS.318.1005G}.  For larger values of $|\psi|$ the theory
developed in Appendix~\ref{s:nonkeplerian} becomes inaccurate.

\begin{figure*}
\centerline{\epsfysize7.5cm\epsfbox{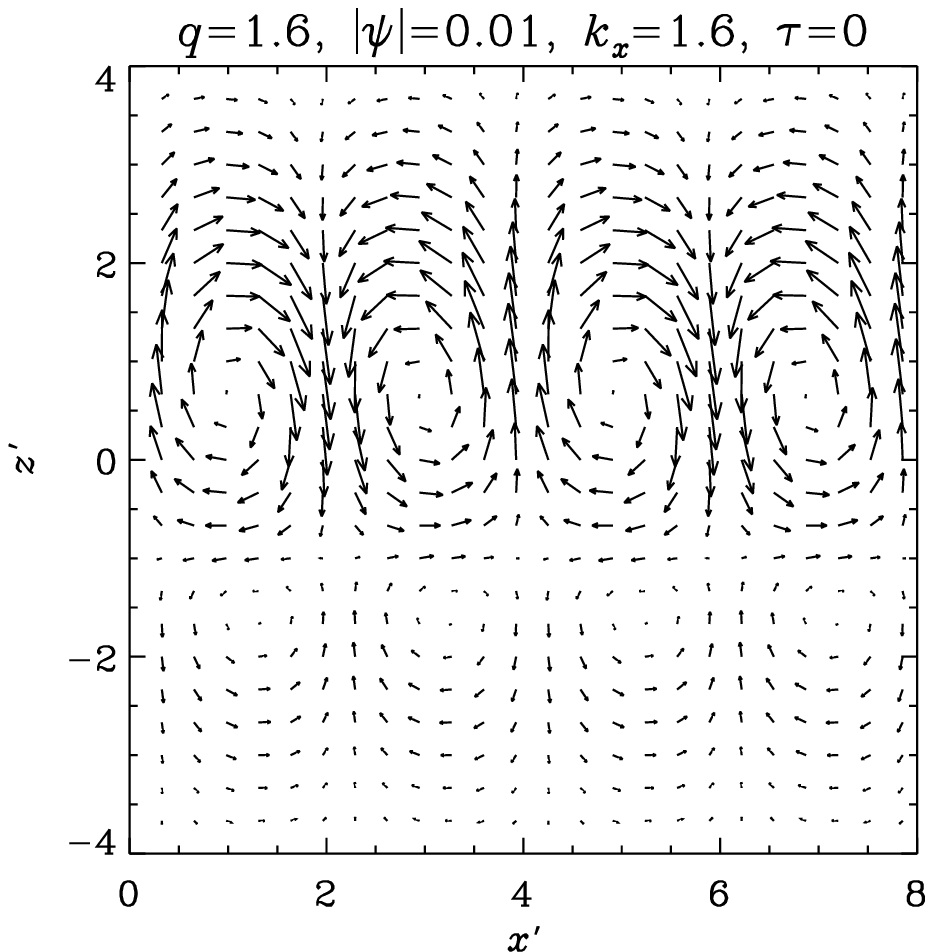}\epsfysize7.5cm\epsfbox{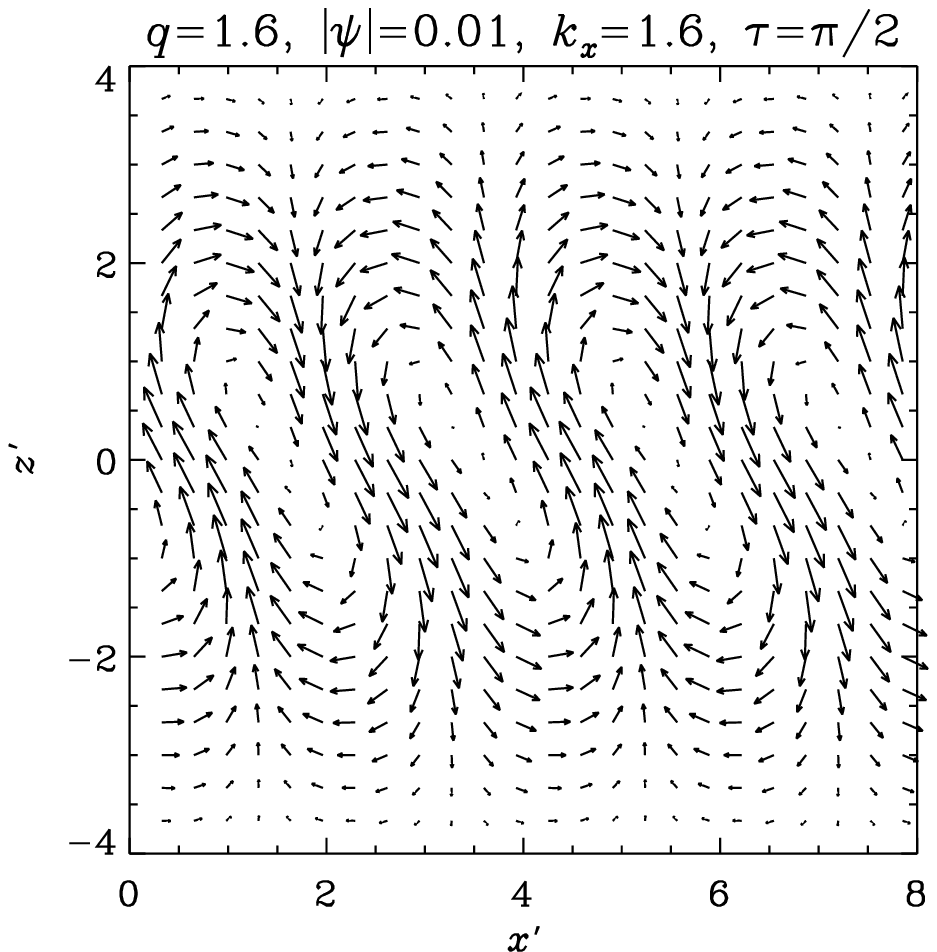}}
\centerline{\epsfysize7.5cm\epsfbox{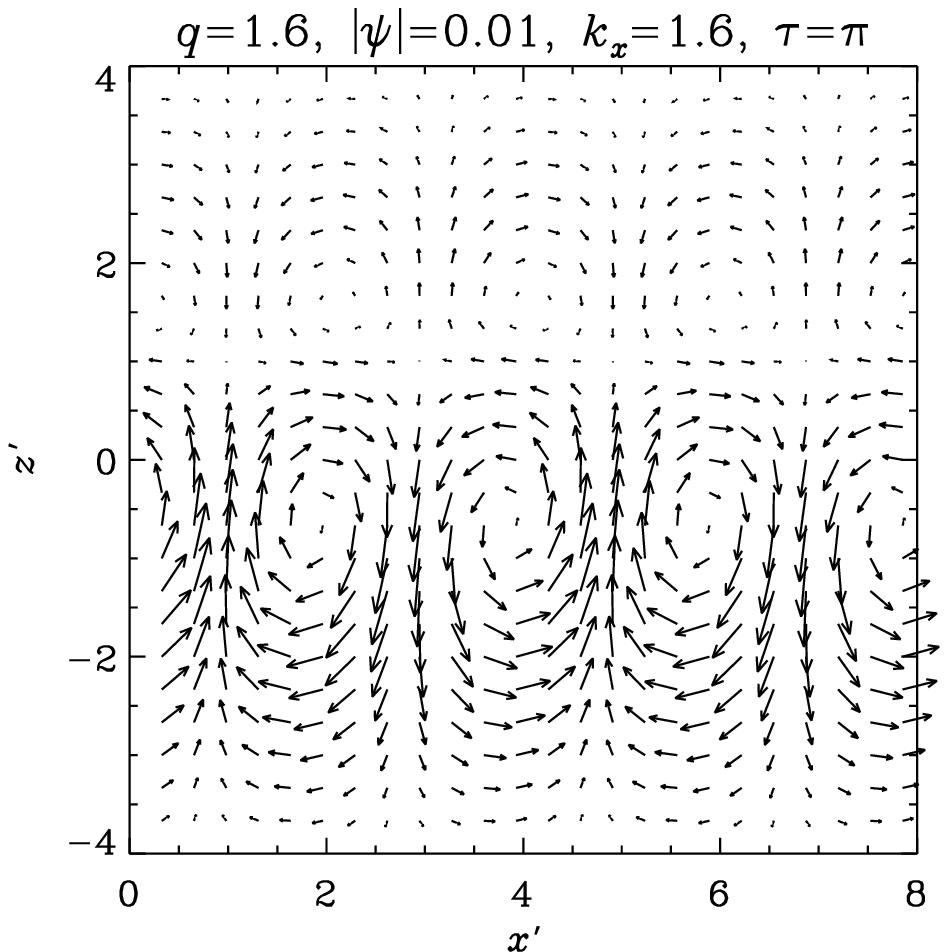}\epsfysize7.5cm\epsfbox{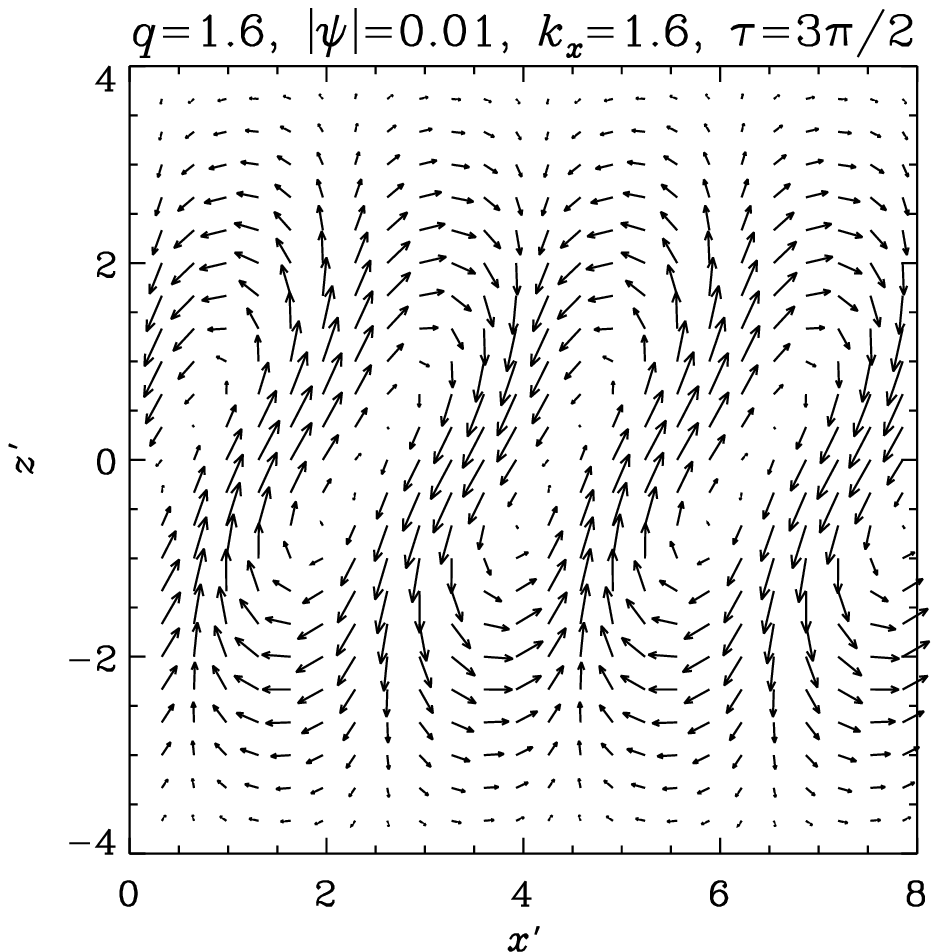}}
\caption{Example of an inviscid growing eigenmode, within the
  first band of parametric instability in the top row of
  Fig.~\ref{f:stability_q=1.6}.  Velocity vectors $\delta\bv$,
  multiplied by $\rho^{1/2}$ to show the distribution of wave energy,
  are shown in the $x'z'$~plane at over two radial wavelengths, at
  orbital phases $\tau=0$, $\pi/2$, $\pi$ and $3\pi/2$.}
\label{f:eigenmode}
\end{figure*}

In Fig.~\ref{f:eigenmode} we show an example of a growing eigenmode,
within the first band of parametric instability in the top row of
Fig.~\ref{f:stability_q=1.6}.  This mode consists mainly of a
superposition of the $n=1$ and $n=2$ inertial waves, coupled
resonantly by the warp so as to cause exponential growth.  The
resulting motion is neither symmetric nor antisymmetric about the
midplane, but involves the oblique motions typical of inertial waves
below the epicyclic frequency (also seen in fig.~1 of
\citealt{2000MNRAS.318.1005G}).  The variation of the velocity field
with orbital phase is due to the interference of the two inertial
waves with different phase speeds $\omega/k_x$; the pattern is not
significantly distorted by shear, because both the warp amplitude and
that of the laminar flow are quite small.  By referring to
equation~(\ref{perturbation_energy}) and to the first panel of fig.~6
of Paper~I (which shows that $u$ is approximately proportional to
$\sin\tau$) we can understand how the mode grows: $\delta v_x$ and
$\delta v_z$ are negatively correlated at phase $\tau=\pi/2$ (panel~2)
where the shear rate $|\psi|\sin\tau+u$ reaches its positive maximum,
and are positively correlated at phase $\tau=3\pi/2$ (panel~4) where
the shear rate reaches its negative minimum.  The Reynolds stresses
associated with this mode are therefore able to release energy from
both the warped orbital motion and the laminar flow.

The situation for a Keplerian disc is more complicated.  Viscosity
must be included in order to obtain laminar flow solutions, and the
analysis of parametric instability is modified, mainly because of the
different phase relationships in the laminar flows.
Fig.~\ref{f:stability_q=1.5} shows some results for very small warps
and Fig.~\ref{f:stability_q=1.5_bis} for larger, but still probably
unobservable, warps.  The viscosity parameter $\alpha$ is chosen in
each case to be sufficiently large that the laminar flow has a modest
amplitude $(|u|<1)$.  Distinct bands of parametric instability are
clearly visible only for very small $|\psi|$ and~$\alpha$, as in the
left panel of Fig.~\ref{f:stability_q=1.5}, where good agreement is
found with the analytical calculation in Appendix~\ref{s:keplerian}.
Note that viscous damping reduces the growth rate at larger values
of~$k_x$.  For larger warp amplitudes, the analytical theory becomes
inaccurate and instability occurs in broader ranges of $k_x$.

\begin{figure*}
\centerline{\epsfysize7.5cm\epsfbox{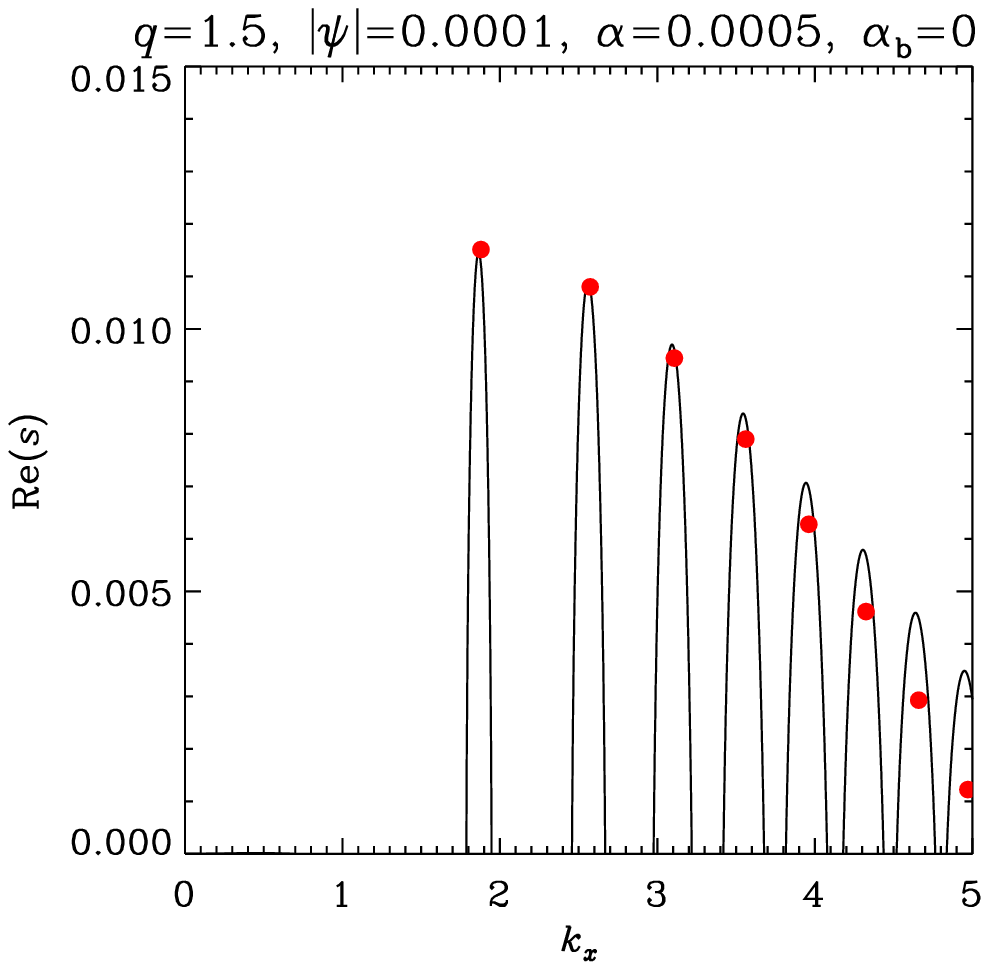}\qquad\epsfysize7.5cm\epsfbox{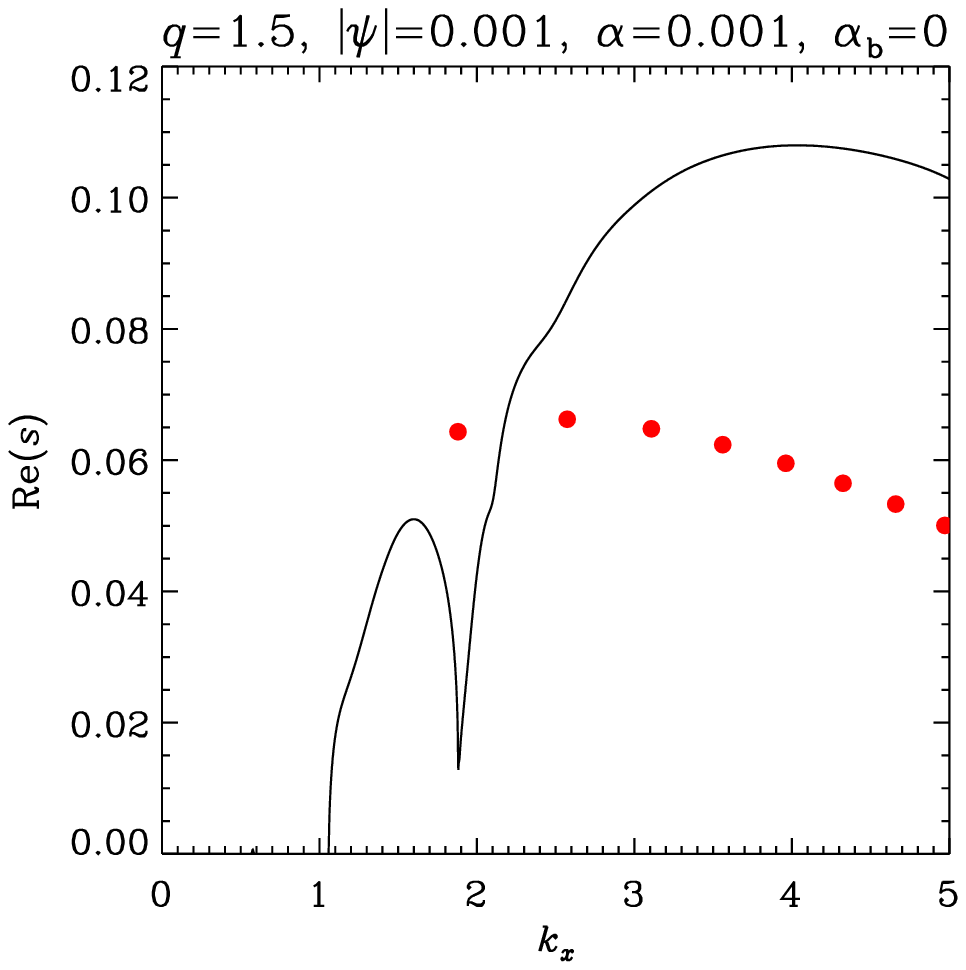}}
\caption{Linear stability of laminar flows for $q=1.5$, including
  viscosity.  The left panel is computed with $N=20$ and the right
  panel with $N=40$, which are sufficient to provide converged results
  for this range of~$k_x$.  The red points show the growth rate at the
  centre of each band of parametric instability, based on the analysis
  in Appendix~\ref{s:keplerian}, and taking into account viscous
  damping; this analysis is inaccurate for the right panel.  Larger
  growth rates are obtained where the resonance bands overlap.}
\label{f:stability_q=1.5}
\end{figure*}

\begin{figure*}
\centerline{\epsfysize7.5cm\epsfbox{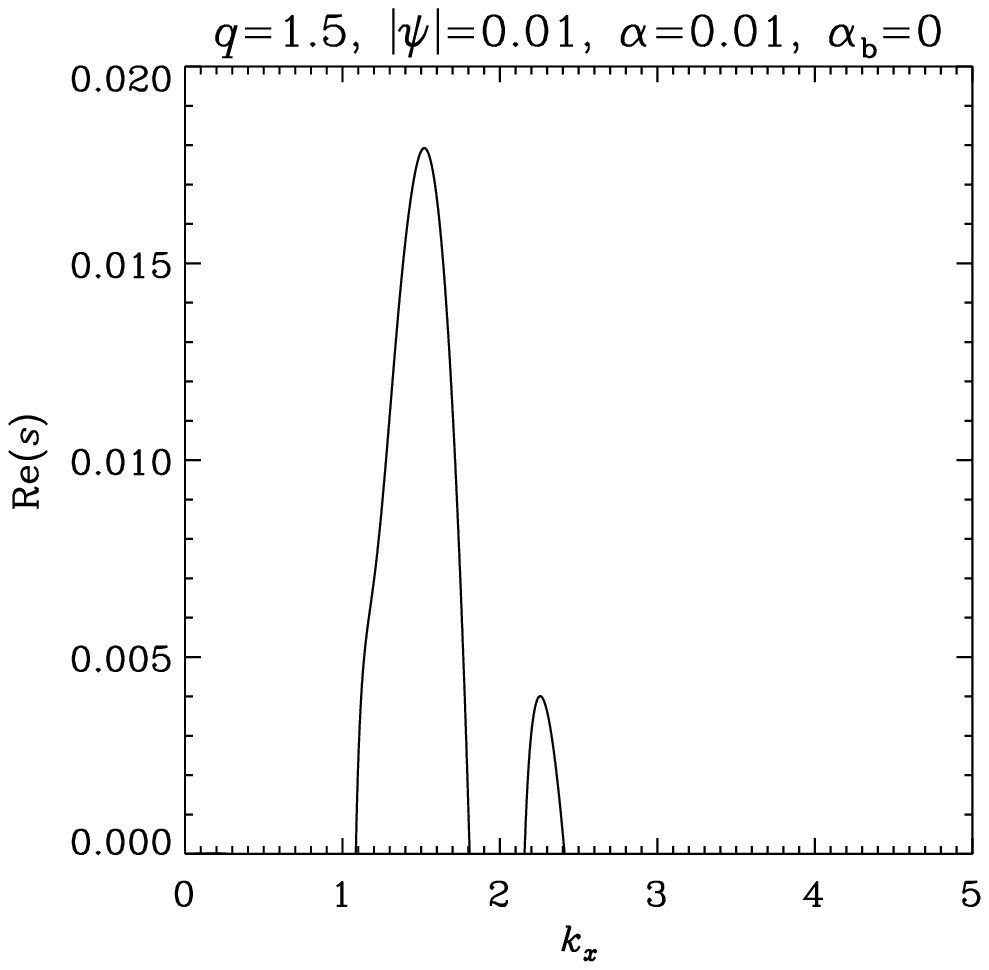}\qquad\epsfysize7.5cm\epsfbox{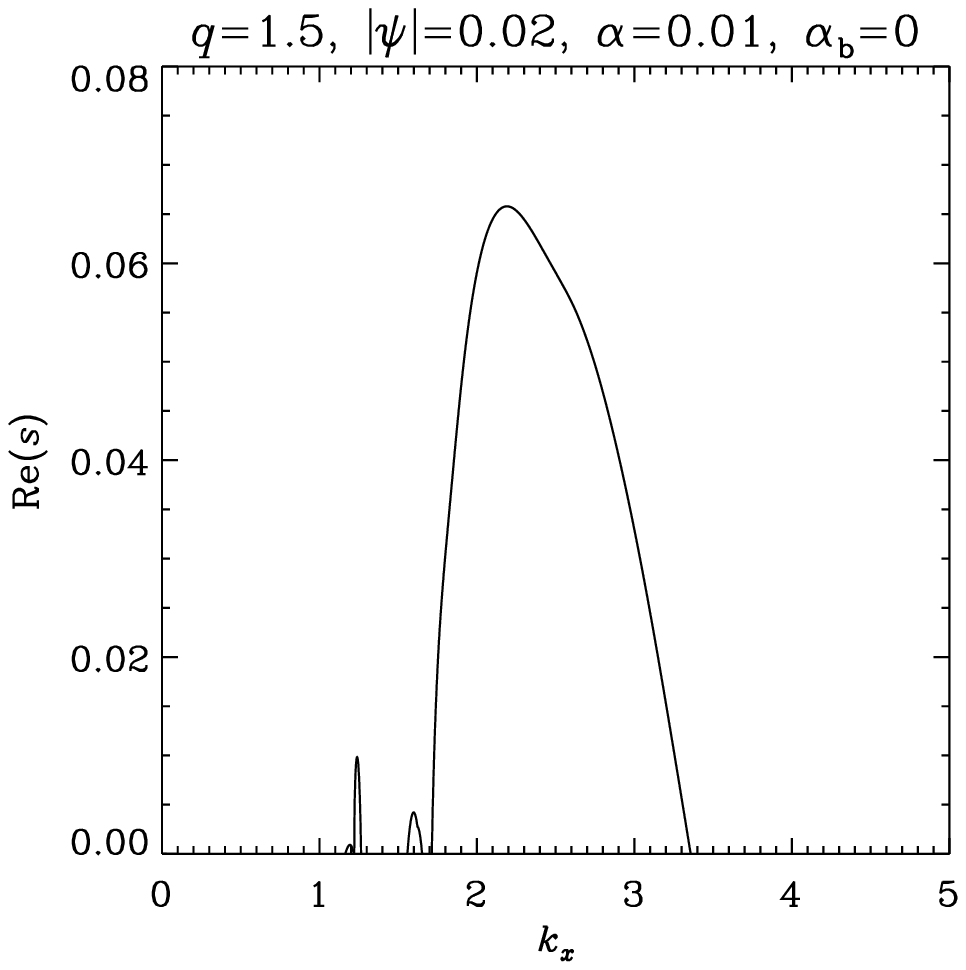}}
\caption{Continuation of Fig.~\ref{f:stability_q=1.5_bis} to larger
  warp amplitudes.  $N=30$ is sufficient for the left panel and $N=80$
  for the right panel.}
\label{f:stability_q=1.5_bis}
\end{figure*}

In situations where the laminar flows have large amplitude ($|u|\ga1$)
and the viscosity is small ($\alpha\ll1$), large growth rates
($\real(s)\ga1$) may be expected.  However, in this regime the
eigenvalues returned by our numerical method do not converge as the
truncation order $N$ is increased, and therefore we do not present any
such results in this paper.  Initially the difficulty takes the form
of spurious modes, presumably arising from the truncation of the
Hermite basis, with non-convergent growth rates that can exceed those
of the genuine modes.  Such unwanted solutions can be excluded by
comparing the eigenvalues calculated with different truncation orders.
Later, however, the entire spectrum fails to converge.  The reason for
this behaviour is probably that the coupling terms proportional to
$\rmi k_x u$ in equations (\ref{dun})--(\ref{dhn}) are strong, and
indeed the problem is worse for larger values of $k_x$.  It may be
simply that the Hermite basis, which is designed for wave modes in an
unwarped disc, is incapable of handling the modes of a warped disc
with strong laminar flows.

It is also possible, however, that modal solutions cease to exist when
the laminar flows are too strong, or when the radial wavenumber is too
large.  After all, it is not obvious why an unwarped isothermal disc
supports acoustic and inertial waves whose energy is confined near the
midplane.  If the vertical gravitational acceleration were constant,
instead of being proportional to $z$, there would be a continuous
spectrum instead of confined modes.  Similarly, in a stably stratified
vertical isothermal disc with adiabatic exponent $\gamma>1$, the
internal gravity waves are not confined but form a continuous
spectrum.  In a shearing sheet without radial structure or boundaries,
non-axisymmetric eigenmodes do not exist because of the inexorable
shear, and are replaced by non-modal solutions.  It is possible that
the oscillatory shear of a strong laminar flow in a warped disc may
prevent the formation of axisymmetric eigenmodes.

So far we have discussed the main bands of instability that occur for
radial wavenumbers $k_x\ga1$.  There are, however, weaker forms of
instability at longer wavelengths.  Close to $k_x=0.925$ in
Fig.~\ref{f:stability_q=1.6}, for example, is a band of instability
(most easily seen in the bottom left panel) whose height and width
scale with $|\psi|^2$ rather than $|\psi|$, indicative of a weaker,
higher-order mode coupling.  We also find viscous overstability at
smaller values of~$k_x$.  Viscous overstability
(\citealt{1978MNRAS.185..629K}; see also
\citealt{2006MNRAS.372.1829L}) tends to cause a slow growth of long
`density' waves in a viscous disc.  In the absence of a warp, the
$n=0$ mode has a growth rate scaling as $\alpha k_x^2$ for $k_x\ll1$,
although this can be suppressed by the addition of a sufficiently
large bulk viscosity.  Modes with $n\ge1$ are not unstable.
Overstability appears in our calculations for viscous warped discs,
although it is not visible in Figs~\ref{f:stability_q=1.5}
and~\ref{f:stability_q=1.5_bis} because the growth rates are too small
compared to the parametric instability.  In the absence of a warp,
overstability occurs at $k_x\ll1$ in an isothermal disc when
$\alpha_\rmb<{\textstyle\f{2}{3}}\alpha$ in the case $q=1.5$, or when
$\alpha_\rmb<{\textstyle\f{5}{3}}\alpha$ in the case $q=1.6$.  This
illustrates the fact that overstability is more likely closer to a
black hole, where (effectively) $q$ is closer to~$2$.  Our
calculations suggest that the presence of a warp and fast laminar
flows acts against overstability.  For example, when $q=1.5$ and
$\alpha_\rmb=0$, overstability is present at $|\psi|=0$ but is
suppressed for $|\psi|\ga1.1\alpha$.

\begin{figure}
\centerline{\epsfysize8cm\epsfbox{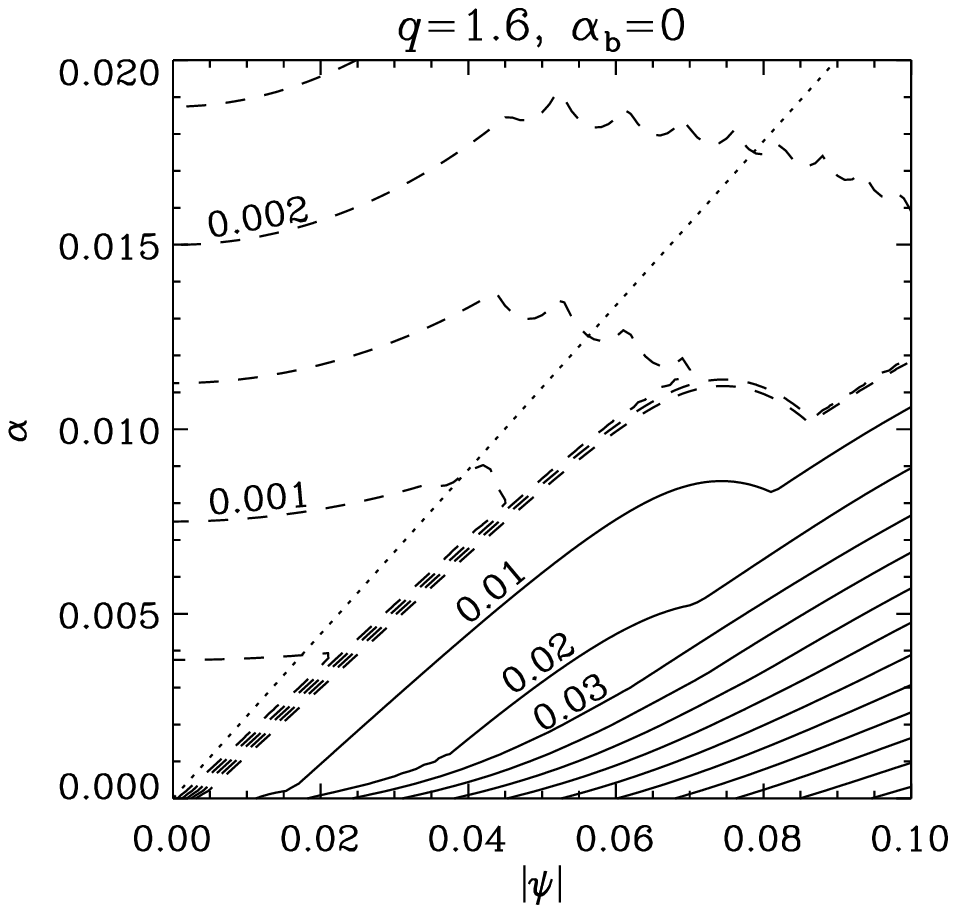}}
\caption{Contours of growth rate, maximized over $0<k_x<5$, for
  $q=1.6$ and $\alpha_\rmb=0$.  Dashed contours: 0.0005, 0.001,
  0.0015, 0.002, 0.0025, showing the region of viscous overstability.
  Solid contours: 0.01, 0.02, 0.03, \dots, 0.12, showing the region of
  parametric instability.  Selected contour values are marked.  Dotted
  line: $\alpha=0.2226|\psi|$, the analytical prediction for the onset
  of parametric instability for $|\psi|\ll1$.}
\label{f:contour_q=1.6}
\end{figure}

\begin{figure}
\centerline{\epsfysize8cm\epsfbox{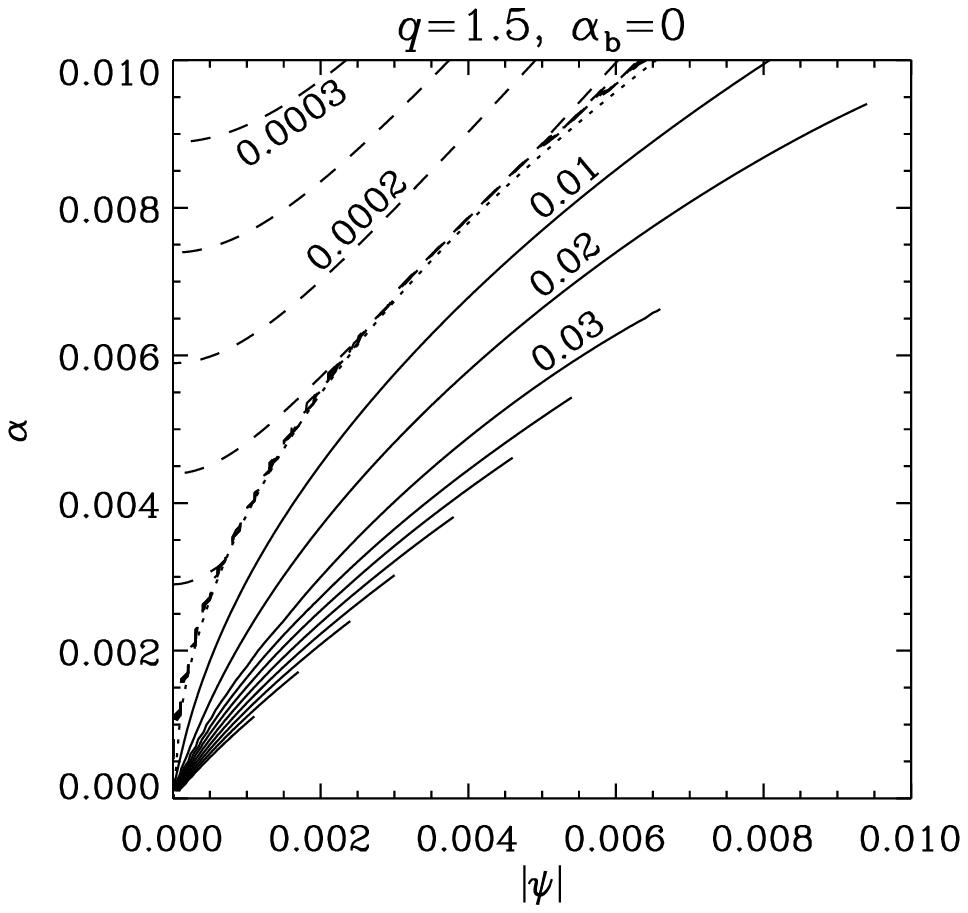}}
\caption{Contours of growth rate, maximized over $0<k_x<5$, for
  $q=1.5$ and $\alpha_\rmb=0$.  Dashed contours: 0.00005, 0.0001,
  0.00015, \dots, 0.0003, showing the region of viscous overstability.
  Solid contours: 0.01, 0.02, 0.03, \dots, 0.1, showing the region of
  parametric instability.  Selected contour values are marked.  Dotted
  line: $\alpha=0.1236|\psi|^{1/2}$, the analytical prediction for the
  onset of parametric instability for $|\psi|\ll1$.  The region of
  fast laminar flows in the lower right part of the diagram is
  excluded from the calculation.}
\label{f:contour_q=1.5}
\end{figure}

We summarize our numerical results in Figs~\ref{f:contour_q=1.6}
and~\ref{f:contour_q=1.5}, where we plot contour lines of the maximum
growth rate obtained for $0<k_x<5$ in the parameter space of warp
amplitude ($|\psi|$) and viscosity ($\alpha$), for non-Keplerian
($q=1.6$) and Keplerian ($q=1.5$) discs.  In the non-Keplerian case
(Fig.~\ref{f:contour_q=1.6}) parametric instability is found for
$\alpha<0.2226|\psi|$ in the limit $|\psi|\ll1$, as determined in
Appendix~\ref{s:nonkeplerian} and indicated by the dotted straight
line.  For larger $|\psi|$ the onset of parametric instability
deviates somewhat from this straight line.  The weaker form of
instability found in most of the remainder of the diagram is the
viscous overstability, which favours longer wavelengths.
Fig.~\ref{f:contour_q=1.5} for the Keplerian case covers a smaller
range of warp amplitudes.  Parametric instability is found for
$\alpha<0.1236|\psi|^{1/2}$ in the limit $|\psi|\ll1$, as determined
in Appendix~\ref{s:keplerian} and indicated by the dotted parabola.
Again, viscous overstability provides a weak growth of longer waves
above this boundary.

In the lower right corner of Fig.~\ref{f:contour_q=1.5}, the strength
of the laminar flows means that growth rates larger than $0.1$ are
expected, but we have not computed this region systematically because
of the onset of a lack of numerical convergence as described above.
For similar reasons we do not extend the contour plots to larger
values of~$|\psi|$ and~$\alpha$.

\section{Discussion}

To the best of our knowledge, the linear hydrodynamic instability of
warped discs that we have described has not been seen in any of the
global, three-dimensional numerical simulations of warped discs.  This
is probably because the simulations have insufficient spatial
resolution or too much viscosity (explicit or numerical) to allow
instability to develop.  For example, the high-resolution SPH
simulations of \citet{2010MNRAS.405.1212L}, which appear compatible
with the laminar nonlinear theory of \citet{1999MNRAS.304..557O}, have
estimated shear viscosities in the range $0.04\la\alpha\la0.5$, in
addition to bulk viscosity.  Their largest (initial) values of
$|\psi|$ are $0.026$ for their low-amplitude warp and $1.35$ for their
high-amplitude warp.  Although we would expect instability to occur
when the amplitude is high and the viscosity is low, it is not clear
whether the unstable modes can be meaningfully resolved, or escape
catastrophic damping, when the SPH smoothing length is typically
$0.6H$.  Furthermore, it is in this regime that the warp tends to
steepen into a `break'.  The high-resolution grid-based simulations of
\citet{2010A&A...511A..77F} have a grid spacing of $0.2H$ in the
vertical direction and an explicit shear viscosity in the range
$0.005\le\alpha\le0.1$.  The warps are mostly very small, however.
The most warped disc they simulate is also the most viscous, and the
instability is probably absent for that reason.  It should also be
borne in mind that global numerical simulations may not have been run
for long enough for weak versions of the hydrodynamic instability to
develop.

Given the inherent limitations of current global simulations, it would
make sense to study the nonlinear outcome of the hydrodynamic
instability of warped discs with local numerical simulations in the
warped shearing box as defined in Paper~I.  Such simulations should be
able to determine the more complicated flows that replace the laminar
solutions in the unstable regions of the parameter space, and allow a
calculation of the torques acting in a warped disc in the presence of
the instability.  Here we can only speculate on what the outcome might
be.  Since the instability feeds on the shear in the laminar flows, it
is possible that it will reduce the amplitude of those flows and may
therefore limit the associated torques that determine the evolution of
the warp, as discussed in section~2 of Paper~I.  In this way the rapid
diffusion of warps in Keplerian discs might be ameliorated, allowing
warps to survive more readily.

Although we have considered only axisymmetric instabilities ($k_y=0$)
in this paper, non-axisymmetric shearing waves are also likely to be
amplified, if only transiently.  The nonlinear evolution could be
computed initially with two-dimensional simulations (independent of
$y$), although the behaviour in three dimensions may be different.

It will also be important to determine the interaction of a warp and
its associated hydrodynamic instability with the magnetorotational
instability (MRI) in discs that are sufficiently coupled to a magnetic
field.  This can be done by solving the MHD equations in a warped
shearing box.  If the MRI is sufficiently powerful, it may suppress
the hydrodynamic instability; the presence of a magnetic field will
certainly modify the low-frequency part of the spectrum of waves in
the disc \citep[e.g.][]{1998MNRAS.297..291O}.  However, if the warp
and its associated flows are sufficiently strong, their effects are
bound to dominate over the MRI.  Indeed, \citet{2000MNRAS.318...47T}
were able to make a numerical study of the interaction between the MRI
and the horizontal shearing motions associated with a warp, and found
evidence that the hydrodynamic instability dominates if the shear is
sufficiently strong.

The model we have investigated in this paper is intentionally
simplified by being isothermal.  If the disc is stably stratified,
buoyancy forces will modify the details of the instability by
replacing the inertial waves with inertia-gravity waves.  However,
since buoyancy forces vanish at the midplane, their effect may be
limited.

\section{Conclusion}

Even warps that are too small to be observable may have important
dynamical consequences for astrophysical discs.  The main reason for
this is the coincidence of the orbital and epicyclic frequencies in a
Keplerian disc.  A warp that is stationary, or nearly so, in a
non-rotating frame, implies oscillatory pressure gradients that drive
horizontal shearing motions close to resonance.  The simplest laminar
flow solutions provide a large torque that leads to the anomalously
rapid diffusion or propagation of warps in Keplerian discs that has
been noted in previous work.  However, these flows are often linearly
unstable and will be replaced by more complicated solutions and
associated torques that remain to be computed.  This dynamics will
alter the evolution of warped discs.  It also has the potential to
introduce non-trivial hydrodynamic behaviour in the form of
turbulence, or at least wave activity, in discs with very small warps,
especially if no other form of turbulence is present.  This effect
could be important, for example, in activating the dead zones of
protoplanetary discs if there is any small misalignment in the system
such as that caused by a planet or a distant binary companion on an
inclined orbit.

We have used a local model, the warped shearing sheet, to compute the
laminar flows and to analyse their linear stability in Keplerian and
non-Keplerian warped discs.  Hydrodynamic instability derives from the
parametric resonance of inertial waves and is widespread, especially
when the viscosity is low or the rotation is close to Keplerian.  Our
analysis is closely related to that of \citet{2000MNRAS.318.1005G},
but is more detailed and is based on the nonlinear laminar flow
solutions computed in a warped shearing sheet.

Future work on warped discs ought to include numerical simulations of
the nonlinear outcome of the hydrodynamic instability in the warped
shearing box in two or three dimensions, and also of its interaction
with the magnetorotational instability.  The linear analysis presented
in this paper should provide some initial guidance for this work.

Closely related problems exist for eccentric discs and tidally
distorted discs.  In both cases the non-circular streamlines present
the fluid with an oscillating geometry that leads to the parametric
excitation of inertial waves.  These are versions of the elliptical
instability, well known in fluid dynamics
\citep[e.g.][]{2002AnRFM..34...83K}.  Thus \citet{1993ApJ...406..596G}
analysed the elliptical instability of tidally distorted discs and
\citet{1994ApJ...422..269R} computed its nonlinear evolution in a
two-dimensional local model, while \citet{2005A&A...432..743P}
analysed the elliptical instability of eccentric discs.  In each case
the level of hydrodynamic activity that can be expected is limited by
the ellipticity of the flow.  An important difference in the case of
warped discs is that coincidence of the orbital and epicyclic
frequencies in a Keplerian disc leads to a resonant enhancement of the
internal flows, so that even a very small warp may produce significant
hydrodynamic activity.

\section*{acknowledgments}

This research was supported by STFC.  We are grateful for the
referee's suggestions.

\newpage

\onecolumn

\appendix

\section{Bound on the growth rate}
\label{s:bound}

The right-hand side of equation~(\ref{perturbation_energy}) can be
written as the integral of an Hermitian form:
\begin{equation}
  -\int\rho(\delta\bv)^*\cdot\mathbf{E}(\tau)\,\delta\bv\,\rmd z',
\label{hermitian}
\end{equation}
where $\mathbf{E}(\tau)$ is the (symmetric) rate-of-strain tensor with
(dimensionless) components given by
\begin{equation}
  E_{xx}=|\psi|\cos\tau\,u,\qquad
  E_{yy}=0,\qquad
  E_{zz}=w,
\end{equation}
\begin{equation}
  E_{xy}=E_{yx}=\half(-q+|\psi|\cos\tau\,v),\qquad
  E_{xz}=E_{zx}=\half(|\psi|\sin\tau+|\psi|\cos\tau\,w+u),\qquad
  E_{yz}=E_{zy}=\half v.
\end{equation}
Let the ordered eigenvalues of $\mathbf{E}(\tau)$ be $E_1(\tau)\le
E_2(\tau)\le E_3(\tau)$.  Then we have
$E_1|\delta\bv|^2\le(\delta\bv)^*\cdot\mathbf{E}\,\delta\bv\le
E_3|\delta\bv|^2$ and so the quantity~(\ref{hermitian}) is bounded
above by $-E_1\int\rho|\delta\bv|^2\,\rmd z'$.  It follows from
equation~(\ref{perturbation_energy}) that the instantaneous growth
rate of the amplitude of the perturbation (in the energy norm) is
limited by
\begin{equation}
  \rmd_\tau\ln\left[\int\rho\left(\half|\delta\bv|^2+\half|\delta h|^2\right)\rmd z'\right]^{1/2}\le-E_1(\tau).
\end{equation}
For the axisymmetric waves discussed in Section~\ref{s:presence}, the
Floquet growth rate is therefore bounded by the orbital average,
\begin{equation}
  \real(s)\le-\langle E_1\rangle_\tau.
\end{equation}
In the absence of a warp the eigenvalues of~$\mathbf{E}$ are $\pm q/2$
and~$0$, leading to the familiar bound $\real(s)\le|q|/2$, i.e.\ half
the shear rate.  In the presence of a warp, the eigenvalues do not
have simple expressions, but various algebraic bounds can be placed on
them if desired.  An example of such a bound is
\begin{equation}
  E_1^2\le E_{ij}E_{ij}=(|\psi|\cos\tau\,u)^2+w^2+\half(-q+|\psi|\cos\tau\,v)^2+\half(|\psi|\sin\tau+|\psi|\cos\tau\,w+u)^2+\half v^2.
\end{equation}

\section{Viscous linearized equations}
\label{s:viscous}

When viscosity is included, the linearized equations (\ref{caldvx})--(\ref{cald}) become
\begin{eqnarray}
  \lefteqn{\mathcal{D}\delta v_x+(\delta v_z+|\psi|\cos\tau\,\delta v_x)u-2\delta v_y=-(\rmi k_x+|\psi|\cos\tau\,\p_z')\delta h}\nonumber\\
  &&+[\rmi k_x+|\psi|\cos\tau(\p_z'-gz')][2\alpha(\rmi k_x+|\psi|\cos\tau\,\p_z')\delta v_x+(\alpha_\mathrm{b}-\twothirds\alpha)\Delta']\nonumber\\
  &&+(\rmi k_x+|\psi|\cos\tau\,\p_z')\{[2\alpha|\psi|\cos\tau\,u+(\alpha_\mathrm{b}-\twothirds\alpha)(w+|\psi|\cos\tau\,u)]\delta h\}\nonumber\\
  &&+\rmi k_y\alpha[(\rmi k_x+|\psi|\cos\tau\,\p_z')\delta v_y+\rmi k_y\delta v_x+(-q+|\psi|\cos\tau\,v)\delta h]\nonumber\\
  &&+\alpha(\p_z'-gz')[(\rmi k_x+|\psi|\cos\tau\,\p_z')\delta v_z+\p_z'\delta v_x]+\alpha(|\psi|\sin\tau+|\psi|\cos\tau\,w+u)\p_z'\delta h,
\end{eqnarray}
\begin{eqnarray}
  \lefteqn{\mathcal{D}\delta v_y+(\delta v_z+|\psi|\cos\tau\,\delta v_x)v+(2-q)\delta v_x=-\rmi k_y\delta h}&\nonumber\\
  &&+[\rmi k_x+|\psi|\cos\tau(\p_z'-gz')]\{\alpha[(\rmi k_x+|\psi|\cos\tau\,\p_z')\delta v_y+\rmi k_y\delta v_x]\}\nonumber\\
  &&+(\rmi k_x+|\psi|\cos\tau\,\p_z')[\alpha(-q+|\psi|\cos\tau\,v)\delta h]\nonumber\\
  &&+\rmi k_y\{2\alpha\,\rmi k_y\delta v_y+(\alpha_\mathrm{b}-\twothirds\alpha)[\Delta'+(w+|\psi|\cos\tau\,u)\delta h]\}\nonumber\\
  &&+(\p_z'-gz')[\alpha(\rmi k_y \delta v_z+\p_z'\delta v_y)]+\alpha v\p_z'\delta h,
\end{eqnarray}
\begin{eqnarray}
  \lefteqn{\mathcal{D}\delta v_z+(\delta v_z+|\psi|\cos\tau\,\delta v_x)w+|\psi|\sin\tau\,\delta v_x=-\p_z'\delta h}\nonumber\\
  &&+[\rmi k_x+|\psi|\cos\tau(\p_z'-gz')]\{\alpha[(\rmi k_x+|\psi|\cos\tau\,\p_z')\delta v_z+\p_z'\delta v_x]\}\nonumber\\
  &&+(\rmi k_x+|\psi|\cos\tau\,\p_z')[\alpha(|\psi|\sin\tau+|\psi|\cos\tau\,w+u)\delta h]+\rmi k_y\alpha(\rmi k_y\delta v_z+\p_z'\delta v_y+v\delta h)\nonumber\\
  &&+(\p_z'-gz')[2\alpha\p_z'\delta v_z+(\alpha_\mathrm{b}-\twothirds\alpha)\Delta']+[2\alpha w+(\alpha_\mathrm{b}-\twothirds\alpha)(w+|\psi|\cos\tau\,u)]\p_z'\delta h,
\end{eqnarray}
\begin{equation}
  \mathcal{D}\delta h-(\delta v_z+|\psi|\cos\tau\,\delta v_x)gz'=-\Delta',
\end{equation}
with
\begin{equation}
  \mathcal{D}=\p_\tau+\rmi(k_xu+k_yv)z'+(w+|\psi|\cos\tau\,u)z'\p_z'
\end{equation}
and
\begin{equation}
  \Delta'=(\rmi k_x+|\psi|\cos\tau\,\p_z')\delta v_x+\rmi k_y\delta v_y+\p_z'\delta v_z.
\end{equation}
When they are projected on to the basis of Hermite polynomials, they become
\begin{eqnarray}
  \lefteqn{\rmd_\tau u_n+\rmi(k_xu+k_yv)[u_{n-1}+(n+1)u_{n+1}]+(w+|\psi|\cos\tau\,u)[nu_n+(n+1)(n+2)u_{n+2}]}&\nonumber\\
  &&+u(w_{n+1}+|\psi|\cos\tau\,u_n)-2v_n=-\rmi k_xh_n-|\psi|\cos\tau\,(n+1)h_{n+1}\nonumber\\
  &&+\rmi k_x\{2\alpha[\rmi k_xu_n+|\psi|\cos\tau(n+1)u_{n+1}]+(\alpha_\mathrm{b}-\twothirds\alpha)\Delta_n\}\nonumber\\
  &&-|\psi|\cos\tau(g-1)(n+1)\{2\alpha[\rmi k_xu_{n+1}+|\psi|\cos\tau(n+2)u_{n+2}]+(\alpha_\mathrm{b}-\twothirds\alpha)\Delta_{n+1}\}\nonumber\\
  &&-|\psi|\cos\tau\,g\{2\alpha[\rmi k_xu_{n-1}+|\psi|\cos\tau\, nu_n]+(\alpha_\mathrm{b}-\twothirds\alpha)\Delta_{n-1}\}\nonumber\\
  &&+[2\alpha|\psi|\cos\tau\,u+(\alpha_\mathrm{b}-\twothirds\alpha)(w+|\psi|\cos\tau\,u)][\rmi k_xh_n+|\psi|\cos\tau(n+1)h_{n+1}]\nonumber\\
  &&+\rmi k_y\alpha[\rmi k_xv_n+|\psi|\cos\tau(n+1)v_{n+1}+\rmi k_yu_n+(-q+|\psi|\cos\tau\,v)h_n]\nonumber\\
  &&-\alpha(g-1)(n+1)[\rmi k_xw_{n+2}+|\psi|\cos\tau(n+2)w_{n+3}+(n+2)u_{n+2}]\nonumber\\
  &&-\alpha g(\rmi k_xw_n+|\psi|\cos\tau\,nw_{n+1}+nu_n)+\alpha(|\psi|\sin\tau+|\psi|\cos\tau\,w+u)(n+1)h_{n+1},
\end{eqnarray}
\begin{eqnarray}
  \lefteqn{\rmd_\tau v_n+\rmi(k_xu+k_yv)[v_{n-1}+(n+1)v_{n+1}]+(w+|\psi|\cos\tau\,u)[nv_n+(n+1)(n+2)v_{n+2}]}&\nonumber\\
  &&+v(w_{n+1}+|\psi|\cos\tau\,u_n)+(2-q)u_n=-\rmi k_yh_n\nonumber\\
  &&+\rmi k_x\alpha[\rmi k_xv_n+|\psi|\cos\tau(n+1)v_{n+1}+\rmi k_yu_n]\nonumber\\
  &&-|\psi|\cos\tau(g-1)(n+1)\alpha[\rmi k_xv_{n+1}+|\psi|\cos\tau(n+2)v_{n+2}+\rmi k_yu_{n+1}]\nonumber\\
  &&-|\psi|\cos\tau\,g\alpha[\rmi k_xv_{n-1}+|\psi|\cos\tau\, nv_n+\rmi k_yu_{n-1}]\nonumber\\
  &&+\alpha(-q+|\psi|\cos\tau\,v)[\rmi k_xh_n+|\psi|\cos\tau(n+1)h_{n+1}]\nonumber\\
  &&+\rmi k_y\{2\alpha\,\rmi k_yv_n+(\alpha_\mathrm{b}-\twothirds\alpha)[\Delta_n+(w+|\psi|\cos\tau\,u)h_n]\}\nonumber\\
  &&-(g-1)(n+1)\alpha[\rmi k_yw_{n+2}+(n+2)v_{n+2}]-g\alpha(\rmi k_yw_n+nv_n)+\alpha v(n+1)h_{n+1},
\end{eqnarray}
\begin{eqnarray}
  \lefteqn{\rmd_\tau w_n+\rmi(k_xu+k_yv)(w_{n-1}+nw_{n+1})+(w+|\psi|\cos\tau\,u)[(n-1)w_n+n(n+1)w_{n+2}]}&\nonumber\\
  &&+w(w_n+|\psi|\cos\tau\,u_{n-1})+|\psi|\sin\tau\,u_{n-1}=-nh_n\nonumber\\
  &&+\rmi k_x\alpha[\rmi k_xw_n+|\psi|\cos\tau\,nw_{n+1}+nu_n]\nonumber\\
  &&-|\psi|\cos\tau(g-1)n\alpha[\rmi k_xw_{n+1}+|\psi|\cos\tau(n+1)w_{n+2}+(n+1)u_{n+1}]\nonumber\\
  &&-|\psi|\cos\tau\,g\alpha[\rmi k_xw_{n-1}+|\psi|\cos\tau(n-1)w_n+(n-1)u_{n-1}]\nonumber\\
  &&+\alpha(|\psi|\sin\tau+|\psi|\cos\tau\,w+u)[\rmi k_xh_{n-1}+|\psi|\cos\tau\,nh_n]\nonumber\\
  &&+\rmi k_y\alpha(\rmi k_yw_n+nv_n+vh_{n-1})-(g-1)n[2\alpha(n+1)w_{n+2}+(\alpha_\mathrm{b}-\twothirds\alpha)\Delta_n]\nonumber\\
  &&-g[2\alpha(n-1)w_n+(\alpha_\mathrm{b}-\twothirds\alpha)\Delta_{n-2}]+[2\alpha w+(\alpha_\mathrm{b}-\twothirds\alpha)(w+|\psi|\cos\tau\,u)]nh_n,
\end{eqnarray}
\begin{eqnarray}
  \lefteqn{\rmd_\tau h_n+\rmi(k_xu+k_yv)[h_{n-1}+(n+1)h_{n+1}]+(w+|\psi|\cos\tau\,u)[nh_n+(n+1)(n+2)h_{n+2}]}&\nonumber\\
  &&-g\{w_n+(n+1)w_{n+2}+|\psi|\cos\tau[u_{n-1}+(n+1)u_{n+1}]\}=-\Delta_n,
\end{eqnarray}
with
\begin{equation}
  \Delta_n=\rmi k_xu_n+|\psi|\cos\tau(n+1)u_{n+1}+\rmi k_yv_n+(n+1)w_{n+2}.
\end{equation}

\section{Parametric instability}
\label{s:parametric}

\subsection{Non-Keplerian case}
\label{s:nonkeplerian}

Parametric instability of a slightly warped disc occurs when the warp
provides a weak nonlinear coupling between wave modes of an unwarped
disc that satisfy an appropriate resonance condition
\citep{2000MNRAS.318.1005G}.  This is a special case of three-wave
coupling in which one of the waves, namely the warp, is of much larger
scale and is therefore effectively of zero wavenumber.  A small amount
of viscous damping of the waves can be allowed for, to compete with
the growth provided by the parametric resonance.  To analyse this
regime we consider an expansion of the equations for small $|\psi|$
and small $\alpha$, letting $\alpha=|\psi|\alpha_1$ and
$\alpha_\rmb=|\psi|\alpha_{\rmb1}$, where $\alpha_1$ and
$\alpha_{\rmb1}$ are of order unity in the limit $|\psi|\ll1$.  To
avoid the additional complications of the coincidence of the orbital
and epicyclic frequencies in a Keplerian disc, we assume here that
$q\ne{\textstyle\f{3}{2}}$.  The Keplerian case is discussed in
Section~\ref{s:keplerian} below.

The laminar flow solutions have the expansion
\begin{equation}
  u=|\psi|S\sin\tau+O(|\psi|^2),\qquad
  v=|\psi|C\cos\tau+O(|\psi|^2),\qquad
  w=O(|\psi|^2),\qquad
  g-1=O(|\psi|^2),
\end{equation}
with
\begin{equation}
  S=\f{1}{2q-3},\qquad
  C=\f{2-q}{2q-3}.
\end{equation}
Then the linearized solutions have the expansion
\begin{equation}
  u_n(\tau)=u_n^{(0)}(\tau_0,\tau_1,\cdots)+|\psi|u_n^{(1)}(\tau_0,\tau_1,\cdots)+O(|\psi|^2),
\end{equation}
and similarly for $v_n$, $w_n$ and $h_n$, where
$(\tau_0,\tau_1,\cdots)=(\tau,|\psi|\tau,\cdots)$ are multiple
time-scales.  Accordingly, the time-derivative $\rmd_\tau$ becomes
$\p_0+|\psi|\p_1+\cdots$, where $\p_i=\p/\p\tau_i$.  The reason for
allowing for multiple time-scales is that, in the regime of interest,
the growth due to parametric resonance and the damping due to
viscosity are weak and occur on a time-scale that is long, by a factor
$O(|\psi|)$, compared to the orbital time-scale.  In this analysis we
are not interested in still longer time-scales and will suppress any
dependence on $\tau_2$, etc.

We consider axisymmetric waves ($k_y=0$), which have a well defined
frequency in the absence of a warp and are therefore suitable for
parametric resonance.  At the leading order, $O(|\psi|^0)$, the
linearized equations yield
\begin{equation}
   L_n\bU_n^{(0)}=\mathbf{0},
\label{leading_order}
\end{equation}
where $L_n$ and $\bU_n^{(0)}$ are a linear operator and a vector of unknowns, given by
\begin{equation}
   L_n=\left[\begin{matrix}\p_0&-2&0&\rmi k_x\\2-q&\p_0&0&0\\0&0&\p_0&n\\\rmi k_x&0&-1&\p_0\end{matrix}\right],\qquad
   \bU_n^{(0)}=\left[\begin{matrix}u_n^{(0)}\\v_n^{(0)}\\w_n^{(0)}\\h_n^{(0)}\end{matrix}\right].
\end{equation}
These are just the linearized equations for axisymmetric waves in an
unwarped disc.  As discussed in Section~\ref{s:absence}, solutions
exist that involve only a single value of~$n$ and are proportional to
$\exp(-\rmi\omega\tau_0)$, where the frequency $\omega$ satisfies the
dispersion relation~(\ref{dr}).  The corresponding eigenvector is
\begin{equation}
  \bU_n^{(0)}=\left[\begin{matrix}\rmi\omega(\omega^2-n)\\(2-q)(\omega^2-n)\\nk_x\omega\\\rmi k_x\omega^2\end{matrix}\right]\rme^{-\rmi\omega\tau_0}.
\end{equation}

At the next order, $O(|\psi|^1)$, the linearized equations yield
\begin{equation}
   L_n\bU_n^{(1)}=\bF_n^{(1)},
\label{next_order}
\end{equation}
where the effective forcing vector is
\begin{equation}
  \bF_n^{(1)}=-\p_1\bU_n^{(0)}+\left[\begin{matrix}-\rmi k_xS\sin\tau[u_{n-1}^{(0)}+(n+1)u_{n+1}^{(0)}]-S\sin\tau\,w_{n+1}^{(0)}-\cos\tau(n+1)h_{n+1}^{(0)}+X_n\\-\rmi k_xS\sin\tau[v_{n-1}^{(0)}+(n+1)v_{n+1}^{(0)}]-C\cos\tau\,w_{n+1}^{(0)}+Y_n\\-\rmi k_xS\sin\tau[w_{n-1}^{(0)}+nw_{n+1}^{(0)}]-\sin\tau\,u_{n-1}^{(0)}+Z_n\\-\rmi k_xS\sin\tau[h_{n-1}^{(0)}+(n+1)h_{n+1}^{(0)}]+\cos\tau\,u_{n-1}^{(0)}\end{matrix}\right]
\end{equation}
and the viscous terms are
\begin{equation}
  X_n=-\alpha_1[(2k_x^2+n)u_n^{(0)}+\rmi k_xw_n^{(0)}]+(\alpha_{\rmb1}-\twothirds\alpha_1)\rmi k_x[\rmi k_xu_n^{(0)}+(n+1)w_{n+2}^{(0)}],
\end{equation}
\begin{equation}
  Y_n=-\alpha_1[(k_x^2+n)v_n^{(0)}+q\,\rmi k_xh_n^{(0)}],
\end{equation}
\begin{equation}
  Z_n=-\alpha_1[(k_x^2+2(n-1))w_n^{(0)}-\rmi k_xnu_n^{(0)}]-(\alpha_{\rmb1}-\twothirds\alpha_1)[\rmi k_xu_{n-2}^{(0)}+(n-1)w_n^{(0)}].
\end{equation}
In these equations, terms due to the warp couple each mode ($n$) with
its neighbours ($n\pm1$).  (The viscous terms also couple $n$ with
$n\pm2$ if $\alpha_\rmb\ne\twothirds\alpha$.)

The operator $L_n$ is singular, because equation~(\ref{leading_order})
has non-trivial solutions.  To discover the associated solvability
conditions, we consider an equation of the form $L_n\bU_n=\bF_n$ with
a forcing vector $\bF_n=[a_n\;b_n\;c_n\;d_n]^\rmT$.  The four
components of this equation can be combined into
\begin{equation}
  \{(\p_0^2+n)[\p_0^2+2(2-q)]+k_x^2\p_0^2\}h_n=-\rmi k_x\p_0(\p_0a_n+2b_n)+[\p_0^2+2(2-q)](c_n+\p_0d_n).
\label{sc}
\end{equation}
For the equation to be solvable, the right-hand side should not
contain any term proportional to $\rme^{-\rmi\omega\tau_0}$, where
$\omega$ is any root of the dispersion relation~(\ref{dr}).  If $a_n$,
$b_n$, $c_n$ and $d_n$ are simply proportional to
$\rme^{-\rmi\omega\tau_0}$ then the solvability condition is
\begin{equation}
  -k_x\omega(-\rmi\omega a_n+2b_n)+[-\omega^2+2(2-q)](c_n-\rmi\omega d_n)=0.
\end{equation}

In order to satisfy the exact conditions for parametric resonance, we
should consider two waves whose frequencies combine (by addition or
subtraction) to give the frequency of the warp, which is equal to~$1$
in our units.  Without loss of generality, we call the frequencies of
the two waves $\omega$ and $\omega+1$.  Their vertical mode numbers
should be neighbouring integers, $m$ and $m+1$, so that they are
coupled by the terms described above.  The exact conditions for
parametric resonance are therefore
\begin{equation}
  (-\omega^2+m)[-\omega^2+2(2-q)]-k_x^2\omega^2=0
\end{equation}
and
\begin{equation}
  (-(\omega+1)^2+m+1)[-(\omega+1)^2+2(2-q)]-k_x^2(\omega+1)^2=0,
\end{equation}
which can be satisfied simultaneously at discrete values of~$k_x$, as
shown in Fig.~\ref{f:resonances}.

Thus we consider a solution in which
\begin{equation}
  \bU_m^{(0)}=\left[\begin{matrix}u_m^{(0)}\\v_m^{(0)}\\w_m^{(0)}\\h_m^{(0)}\end{matrix}\right]=A_m(\tau_1)\left[\begin{matrix}\rmi\omega(\omega^2-m)\\(2-q)(\omega^2-m)\\mk_x\omega\\\rmi k_x\omega^2\end{matrix}\right]\rme^{-\rmi\omega\tau_0}
\label{leading_order_solution1}
\end{equation}
and
\begin{equation}
  \bU_{m+1}^{(0)}=\left[\begin{matrix}u_{m+1}^{(0)}\\v_{m+1}^{(0)}\\w_{m+1}^{(0)}\\h_{m+1}^{(0)}\end{matrix}\right]=A_{m+1}(\tau_1)\left[\begin{matrix}\rmi(\omega+1)((\omega+1)^2-(m+1))\\(2-q)((\omega+1)^2-(m+1))\\(m+1)k_x(\omega+1)\\\rmi k_x(\omega+1)^2\end{matrix}\right]\rme^{-\rmi(\omega+1)\tau_0},
\label{leading_order_solution2}
\end{equation}
while $\bU_n^{(0)}=\mathbf{0}$ for other values of~$n$, where
$A_m(\tau_1)$ and $A_{m+1}(\tau_1)$ are the slowly varying amplitudes
of the two waves.  The solution
(\ref{leading_order_solution1})--(\ref{leading_order_solution2})
satisfies equation~(\ref{leading_order}) because it is a linear
combination of eigenmodes.  When substituted into
equation~(\ref{next_order}) it generates components of the forcing
vector $\bF_n^{(1)}$ with various frequencies.  Many of these terms
produce a non-resonant response $\bU_n^{(1)}$ that could be found by
solving equation~(\ref{next_order}).  Of interest here, however, are
the forcing terms that resonate with the free wave modes.  The part of
$\bF_m^{(1)}$ that is proportional to $\rme^{-\rmi\omega\tau_0}$ and
therefore resonates with mode~$m$ is (omitting viscous terms)
\begin{equation}
  -\p_1\left[\begin{matrix}u_m^{(0)}\\v_m^{(0)}\\w_m^{(0)}\\h_m^{(0)}\end{matrix}\right]+\left[\begin{matrix}-\half k_xS(m+1)u_{m+1}^{(0)}+\half\rmi Sw_{m+1}^{(0)}-\half(m+1)h_{m+1}^{(0)}\\-\half k_xS(m+1)v_{m+1}^{(0)}-\half Cw_{m+1}^{(0)}\\-\half k_xSmw_{m+1}^{(0)}\\-\half k_xS(m+1)h_{m+1}^{(0)}\end{matrix}\right]\rme^{\rmi\tau_0},
\end{equation}
while the part of
$\bF_{m+1}^{(1)}$ that is proportional to $\rme^{-\rmi(\omega+1)\tau_0}$ and
therefore resonates with mode~$m$ is (again omitting viscous terms)
\begin{equation}
  -\p_1\left[\begin{matrix}u_{m+1}^{(0)}\\v_{m+1}^{(0)}\\w_{m+1}^{(0)}\\h_{m+1}^{(0)}\end{matrix}\right]+\left[\begin{matrix}\half k_xSu_m^{(0)}\\\half k_xSv_m^{(0)}\\\half k_xSw_m^{(0)}-\half\rmi u_m^{(0)}\\\half k_xSh_m^{(0)}+\half u_m^{(0)}\end{matrix}\right]\rme^{-\rmi\tau_0}.
\end{equation}
Applying the solvability condition~(\ref{sc}) to each of these forcing
vectors leads to the following relations between the amplitudes of the
two waves:
\begin{equation}
  \p_1A_m=C_1A_{m+1},\qquad
  \p_1A_{m+1}=C_2A_m,
\end{equation}
where
\begin{equation}
   C_1=\f{[\omega^3(\omega+2)(\omega+q-1)-(2-q)(2\omega+1)m](m+1)k_x}{2(2q-3)\omega[-\omega^4+2(2-q)m]},
\end{equation}
\begin{equation}
   C_2=\f{-(2\omega+1)[\omega^2(\omega+1)^2-2(2-q)m]k_x^2-(2q-3)\omega(\omega+2)(\omega^2-m)(\omega^2+2\omega+2q-3)}{4(2q-3)(\omega+1)[-(\omega+1)^4+2(2-q)(m+1)]k_x}.
\end{equation}
The solutions of these coupled equations are proportional to
$\rme^{s_1\tau_1}$, with scaled growth rates $s_1$ given by
$s_1^2=C_1C_2$.  The true growth rate of the parametric instability is
$s=|\psi|s_1$.

We have evaluated the growth rates for all the possible resonances
shown in Fig.~\ref{f:resonances} (right panel).  Those that involve
couplings between an inertial mode and an acoustic mode have
$C_1C_2<0$ and do not lead to instability.  Those that involve
couplings between two inertial modes have $C_1C_2>0$ and do lead to
instability.  The growth rates for $q=1.6$ and $|\psi|=0.01$ or $0.02$
are plotted as points in Fig.~\ref{f:stability_q=1.6} together with
the results of the full numerical calculations.

In the limit $m\gg1$ the following approximations hold:
\begin{equation}
  \omega=-\f{1}{2}+\f{(15-8q)}{64(2-q)}m^{-1}+O(m^{-2}),
\end{equation}
\begin{equation}
  k_x^2=(15-8q)m+\f{15-8q}{4}+O(m^{-1}),
\end{equation}
\begin{equation}
  C_1C_2=\f{(3-q)^2(15-8q)}{1024(2-q)^2(2q-3)^2}+O(m^{-1}).
\end{equation}
Instability is therefore possible in this limit if
$q<{\textstyle\f{15}{8}}$.  For $q=1.6$, the limiting value of the
growth rate is $0.8111|\psi|$.  Note that the numerical growth rates
at large $k_x$ in Fig.~\ref{f:stability_q=1.6} exceed this value
because of the overlap of resonances.

The calculation given above is readily extended to allow for a slight
detuning of the resonance, of order $|\psi|$, and a slight damping of
the waves, also of order $|\psi|$.  This leads to the system of equations
\begin{equation}
  \p_1A_m=C_1A_{m+1}-C_3A_m,\qquad
  \p_1A_{m+1}=C_2A_m-C_4A_{m+1}+\rmi\omega_1 A_{m+1},
\end{equation}
where the viscous damping coefficients are given by
\begin{eqnarray}
  \lefteqn{C_3[-\omega^4+2(2-q)m]=\alpha_1\{\omega^2(-\omega^2+2)k_x^2+\omega^2(-\omega^2+1)m+2(2-q)[(-\omega^2+m)k_x^2+m(m-1)]\}}&\nonumber\\
  &&+\half(\alpha_{\rmb1}-{\textstyle\f{2}{3}}\alpha_1)\{\omega^2[(-\omega^2+m)k_x^2-m(m-1)]+2(2-q)m(m-1)\}
\end{eqnarray}
and a similar expression for $C_4$, but with $m\mapsto m+1$ and
$\omega\mapsto\omega+1$.  Here the detuning (i.e.\ the mismatch
between the frequency difference of the two waves and the
frequency~$1$ of the warp) is $\omega_1|\psi|$.  In this case
instability occurs at the centre of the resonance ($\omega_1=0$) if
$C_1C_2>C_3C_4$.  The half-width of the resonance (in terms of
frequency) can be seen from the condition for instability,
\begin{equation}
  \omega_1^2<\f{(C_3+C_4)^2(C_1C_2-C_3C_4)}{C_3C_4}.
\end{equation}
However, in the inviscid case the width is given by
\begin{equation}
  \omega_1^2<4C_1C_2,
\end{equation}
which means that the half-width of the resonance is twice the growth
rate at the centre of the resonance.  In fact, the inviscid growth
rate anywhere within the resonant width is
$(C_1C_2-{\textstyle\f{1}{4}}\omega_1^2)^{1/2}$.  The viscous and
inviscid expressions for the width agree if
$C_3=C_4\ll(C_1C_2)^{1/2}$.  The half-width of the resonance in terms
of wavenumber can be found by differentiating the dispersion relation:
small departures $\omega_1|\psi|$ and $k_1|\psi|$ from the centre of
the resonance are related by
\begin{equation}
  \f{\omega_1}{k_1}=\f{\omega^3k_x}{[-\omega^4+2(2-q)m]}-\f{(\omega+1)^3k_x}{[-(\omega+1)^4+2(2-q)(m+1)]}.
\end{equation}

For the first successful resonance in the case $q=1.6$, which has
$m=1$, $k_x=1.6038$ and $\omega=-0.4374$, we have $C_1=1.7125$,
$C_2=0.2692$,
$C_3=3.5485\alpha_1+0.2607(\alpha_{\rmb1}-{\textstyle\f{2}{3}}\alpha_1)$,
$C_4=4.5787\alpha_1+0.7792(\alpha_{\rmb1}-{\textstyle\f{2}{3}}\alpha_1)$
and $\omega_1/k_1=-0.3662$.  We then have instability for
$\alpha<0.2044|\psi|$ in the case
$\alpha_\rmb={\textstyle\f{2}{3}}\alpha$, or $\alpha<0.2226|\psi|$ in
the case $\alpha_\rmb=0$.  The wavenumber-half-width of the resonance
is $3.708|\psi|$ in the inviscid case.

\subsection{Keplerian case}
\label{s:keplerian}

In the Keplerian case $q={\textstyle\f{3}{2}}$ we must include
viscosity in order to find laminar flow solutions.  The amplitude of
the laminar flows, and therefore the growth rate of the parametric
instability, scale with $|\psi|/\alpha$, while the viscous damping
rate scales with $\alpha$.  In order to allow these to compete, we let
$\alpha=|\psi|^{1/2}\alpha_1$ and
$\alpha_\rmb=|\psi|^{1/2}\alpha_{\rmb1}$ in this section, where
$\alpha_1$ and $\alpha_{\rmb1}$ are of order unity in the limit
$|\psi|\ll1$.

Now the laminar flow solutions have the expansion
\begin{equation}
  u=|\psi|^{1/2}C\cos\tau+O(|\psi|^2),\qquad
  v=|\psi|^{1/2}S\sin\tau+O(|\psi|^2),\qquad
  w=O(|\psi|^{3/2}),\qquad
  g-1=O(|\psi|^{3/2}),
\end{equation}
with
\begin{equation}
  C=\f{1}{2\alpha_1},\qquad
  S=-\f{1}{4\alpha_1}.
\end{equation}
Their phase relationship to the warp is different from the inviscid
non-Keplerian case because now the epicyclic oscillator is driven at
its natural frequency and the response is limited by viscous damping.
The linearized solutions have the expansion
\begin{equation}
  u_n(\tau)=u_n^{(0)}(\tau_0,\tau_1,\cdots)+|\psi|^{1/2}u_n^{(1)}(\tau_0,\tau_1,\cdots)+O(|\psi|),
\end{equation}
etc., where now
$(\tau_0,\tau_1,\cdots)=(\tau,|\psi|^{1/2}\tau,\cdots)$.  The rest of
the argument goes through similarly to the non-Keplerian case, but
with the simpler forcing vector
\begin{equation}
  \bF_n^{(1)}=-\p_1\bU_n^{(0)}+\left[\begin{matrix}-\rmi k_xC\cos\tau[u_{n-1}^{(0)}+(n+1)u_{n+1}^{(0)}]-C\cos\tau\,w_{n+1}^{(0)}+X_n\\-\rmi k_xC\cos\tau[v_{n-1}^{(0)}+(n+1)v_{n+1}^{(0)}]-S\sin\tau\,w_{n+1}^{(0)}+Y_n\\-\rmi k_xC\cos\tau[w_{n-1}^{(0)}+nw_{n+1}^{(0)}]+Z_n\\-\rmi k_xC\cos\tau[h_{n-1}^{(0)}+(n+1)h_{n+1}^{(0)}]\end{matrix}\right].
\end{equation}
The part of
$\bF_m^{(1)}$ that is proportional to $\rme^{-\rmi\omega\tau_0}$ and
therefore resonates with mode~$m$ is (omitting viscous terms)
\begin{equation}
  -\p_1\left[\begin{matrix}u_m^{(0)}\\v_m^{(0)}\\w_m^{(0)}\\h_m^{(0)}\end{matrix}\right]+\left[\begin{matrix}-\half\rmi k_xC(m+1)u_{m+1}^{(0)}-\half Cw_{m+1}^{(0)}\\-\half\rmi k_xC(m+1)v_{m+1}^{(0)}+\half\rmi Sw_{m+1}^{(0)}\\-\half\rmi k_xCmw_{m+1}^{(0)}\\-\half\rmi k_xC(m+1)h_{m+1}^{(0)}\end{matrix}\right]\rme^{\rmi\tau_0},
\end{equation}
while the part of
$\bF_{m+1}^{(1)}$ that is proportional to $\rme^{-\rmi(\omega+1)\tau_0}$ and
therefore resonates with mode~$m$ is (again omitting viscous terms)
\begin{equation}
  -\p_1\left[\begin{matrix}u_{m+1}^{(0)}\\v_{m+1}^{(0)}\\w_{m+1}^{(0)}\\h_{m+1}^{(0)}\end{matrix}\right]+\left[\begin{matrix}-\half\rmi k_xCu_m^{(0)}\\-\half\rmi k_xCv_m^{(0)}\\-\half\rmi k_xCw_m^{(0)}\\-\half\rmi k_xCh_m^{(0)}\end{matrix}\right]\rme^{-\rmi\tau_0}.
\end{equation}
From the solvability conditions we now find
\begin{equation}
   C_1=\f{(2\omega+1)[\omega^3(\omega+2)-m](m+1)\rmi k_x}{8\alpha_1\omega(-\omega^4+m)},
\end{equation}
\begin{equation}
   C_2=\f{(2\omega+1)[\omega^2(\omega+1)^2-m]\rmi k_x}{8\alpha_1(\omega+1)[-(\omega+1)^4+(m+1)]}.
\end{equation}
Although $C_1$ and $C_2$ are now imaginary, the previous analysis can
still be applied and instability is possible if $C_1C_2>0$, which
again occurs for couplings between two inertial modes.

For the first successful resonance, which has $m=1$, $k_x=1.8795$ and
$\omega=-0.4325$, we have $C_1=0.1712\rmi/\alpha_1$,
$C_2=-0.02769\rmi/\alpha_1$,
$C_3=4.3750\alpha_1+0.2784(\alpha_{\rmb1}-{\textstyle\f{2}{3}}\alpha_1)$,
$C_4=5.4174\alpha_1+0.8608(\alpha_{\rmb1}-{\textstyle\f{2}{3}}\alpha_1)$
and $\omega_1/k_1=-0.3387$.  We then have instability for
$\alpha<0.1189|\psi|^{1/2}$ in the case
$\alpha_\rmb={\textstyle\f{2}{3}}\alpha$, or
$\alpha<0.1236|\psi|^{1/2}$ in the case $\alpha_\rmb=0$.  The
wavenumber-half-width of the resonance is $0.4065|\psi|/\alpha$ in the
limit of negligible damping.

\label{lastpage}

\end{document}